\documentclass[fleqn]{article}
\usepackage{longtable}
\usepackage{graphicx}

\begin {document}
\begin{flushleft}
{\LARGE
{\bf Energy levels, radiative rates and electron impact excitation rates for transitions in He-like Kr XXXV}
}\\

\vspace{1.5 cm}

{\bf {Kanti  M  ~Aggarwal and Francis  P  ~Keenan}}\\ 

\vspace*{1.0cm}

Astrophysics Research Centre, School of Mathematics and Physics, Queen's University Belfast, Belfast BT7 1NN, Northern Ireland, UK\\ 

\vspace*{0.5 cm} 

e-mail: K.Aggarwal@qub.ac.uk \\

\vspace*{1.50cm}

Received  23 April 2012\\
Accepted for publication 27 July 2012 \\
Published xx  Month 2012 \\
Online at stacks.iop.org/PhysScr/vol/number \\

\vspace*{1.5cm}

PACS Ref: 32.70 Cs, 34.80 Dp, 95.30 Ky

\vspace*{1.0 cm}

\hrule

\vspace{0.5 cm}
{\Large {\bf S}} This article has associated online supplementary data files \\
Tables 2 and 4 are available only in the electronic version at stacks.iop.org/PhysScr/vol/number

\end{flushleft}

\clearpage


\begin{abstract}

We report calculations of energy levels, radiative rates and electron impact excitation cross sections and rates for transitions in He-like Kr XXXV.  The
{\sc grasp} (general-purpose relativistic atomic structure package) is adopted for calculating energy levels and radiative rates. For determining the collision strengths and
subsequently the excitation rates, the Dirac Atomic R-matrix Code ({\sc darc}) is used.  Oscillator strengths, radiative rates and line strengths are reported for all E1, E2,
M1, and M2 transitions among the lowest 49 levels. Additionally, theoretical lifetimes are listed for all 49 levels. Collision strengths are averaged over a Maxwellian velocity distribution and the effective collision strengths  obtained listed over a wide temperature range up to 10$^{8.1}$ K. Comparisons are made with similar data obtained with the Flexible Atomic Code ({\sc fac}) to  assess the accuracy of the results and to highlight the importance of resonances, included in calculations with {\sc darc}, in the determination of effective collision strengths. Differences between the collision strengths from {\sc darc} and {\sc fac}, particularly for forbidden transitions, are also discussed.  Finally, discrepancies between the present results of effective collision strengths from the {\sc darc} code and earlier semi-relativistic $R$-matrix data are noted over a wide range of electron temperatures for many transitions of Kr XXXV.

\end{abstract}

\clearpage

\section{Introduction}

Krypton is used in a variety of experiments in tokamak fusion plasmas and has diagnostic  applications. High temperatures in fusion plasmas, in excess of 10$^7$ K,  give rise to many ionisation stages, including He-like Kr XXXV. Since krypton is increasingly being injected in  fusion machines for the diagnostics of plasmas \cite{kdl}, its study has become more important, particularly with the upcoming ITER project. However, to analyse observations, atomic data are required for a variety of parameters, such as energy levels, radiative rates (A- values), and excitation rates or equivalently the effective collision strengths ($\Upsilon$), which are obtained from the electron impact collision strengths ($\Omega$).  With this in view we have already reported atomic data  for B-like and F-like Kr ions \cite{adt1},\cite{adt2},\cite{adt3}, and here we provide similar data for He-like Kr XXXV.

Emission lines of He-like ions have been widely observed from a variety of  laboratory plasmas.  For example, spectra of Kr XXXV have been recorded in a high temperature Z-pinch discharge by Golt's {\em et al} \cite{ya}. Of particular interest are the resonance ($w$: 1s$^2$ $^1$S$_0$ -- 1s2p $^1$P$^{\rm o}_1$), intercombination ($x$ and $y$: 1s$^2$ $^1$S$_0$ -- 1s2p $^3$P$^{\rm o}_{2,1}$), and forbidden ($z$: 1s$^2$ $^1$S$_0 $ -- 1s2s $^3$S$_1$) lines of Kr XXXV \cite{pb1},\cite{pb2}. Considering the importance of Kr XXXV, a few calculations for its atomic parameters have already been performed using a variety of methods.  Saloman \cite{nist} has compiled data from multiple calculations and has critically evaluated  energy levels for many Kr ions, including Kr XXXV, and has  posted data on the NIST (National Institute of Standards and Technology)  website {\tt http://www.nist.gov/pml/data/asd.cfm}. Similarly, wavelengths are available for some transitions on the NIST website, but their quoted accuracy is not high. Recently, Zhang {\em et al} \cite{liz} have reported wavelengths and A- values, but only for the 1s$^2$ $^1$S$_0$ - $^{1,3}$P$^o_1$ transitions of Kr XXXV. Similarly, the collisional atomic data  for Kr XXXV are also very limited. For example, Pindzola and Carter \cite{pc} have calculated collision strengths using the relativistic {\em distorted-wave} (DW) method, but only for transitions within the levels of the $n$=2 configuration, and within a limited energy range of $\sim$ 13--15 keV.  Sampson {\em et al} \cite{dhs} have reported collision strengths, but only for transitions from the lowest three levels to higher excited levels. Their calculations are based on the Coulomb-Born-exchange method and do not include the contribution of resonances, which can be very important as demonstrated in our earlier papers  on other Kr ions \cite{adt2},\cite{adt3} as well as other He-like ions -- see, for example, Aggarwal and Keenan \cite{caxix} and references therein.  Finally, Griffin and Ballance \cite{damp} have recently performed fully relativistic $R$-matrix calculations for all transitions among the lowest 49 fine-structure levels of Kr XXXV, which belong to the 1s$^2$, 1s2$\ell$, 1s3$\ell$, 1s4$\ell$, and 1s5$\ell$ configurations, but have reported results of effective collision strengths only for transitions from the ground level up to 1s4f $^1$F$^o_3$, i.e.  level 31. Therefore, in this paper we report a complete set of results (namely energy levels, radiative rates, lifetimes, and effective collision strengths) for all 1176 transitions among the lowest 49 fine-structure levels of Kr XXXV. Furthermore, we also provide the A- values for four types of transitions, namely electric dipole (E1), electric quadrupole (E2), magnetic dipole (M1), and  magnetic quadrupole (M2), as these are required in  a complete plasma model.

For our calculations we employ the fully relativistic {\sc grasp} (general-purpose relativistic atomic structure  package) code for the determination of wavefunctions, originally developed by Grant {\em et al} \cite{grasp0} and subsequently revised by several workers, under the names {\sc grasp1} \cite{grasp1},  {\sc grasp2} \cite{grasp2}, and {\sc grasp2k}  \cite{grasp2k}. However, the version adopted here is {\sc grasp0}, which is based on \cite{grasp0} and is revised by Dr P H Norrington.  This version contains most of the modifications undertaken in the other revised codes and is available on the website  {\tt http://web.am.qub.ac.uk/DARC/}.  {\sc grasp} is a fully relativistic code, and is based on the $jj$ coupling scheme.  Further relativistic corrections arising from the Breit interaction and QED effects (vacuum polarization and Lamb shift) have also been included. Additionally, we have used the option of {\em extended average level} (EAL),  in which a weighted (proportional to 2$j$+1) trace of the Hamiltonian matrix is minimized. This produces a compromise set of orbitals describing closely lying states with  moderate accuracy. For our calculations of $\Omega$, we have adopted the {\em Dirac Atomic $R$-matrix Code} ({\sc darc}) of P H Norrington and I P Grant ({\tt http://web.am.qub.ac.uk/DARC/}). Finally, for comparison purposes we have performed parallel calculations with the {\em Flexible Atomic Code} ({\sc fac}) of Gu \cite{fac}, available from the website {\tt {\verb+http://sprg.ssl.berkeley.edu/~mfgu/fac/+}}. This is also a fully relativistic code which provides a variety of atomic parameters, and (generally) yields results for energy levels and radiative rates comparable to {\sc grasp}  -- see, for example, \cite{adt1}, \cite{caxix}, \cite{fe15}.  However, differences in collision strengths and subsequently in effective collision strengths with those obtained from {\sc darc} can be large, particularly for  forbidden transitions, as demonstrated in some of our earlier papers \cite{caxix}, and also discussed below in sections 5 and 6. Hence results from {\sc fac} will be helpful in assessing the accuracy of our energy levels and radiative rates, and in estimating the contribution of resonances to the determination of effective collision strengths, included in calculations from {\sc darc} but neglected in {\sc fac}.

\section{Energy levels}

The 1s$^2$, 1s2$\ell$, 1s3$\ell$, 1s4$\ell$, and 1s5$\ell$ configurations of Kr XXXV  give rise to the lowest 49 levels listed in Table 1,  where we compare our level energies from {\sc grasp} (obtained {\em without} and {\em with} the inclusion of Breit and QED effects) with the critically  compiled data by Saloman \cite{nist}. Also included in the table are our results obtained with the {\sc fac} code (FAC1), including the same CI (configuration interaction) as in {\sc grasp}.  Our level energies obtained without the Breit and QED effects (GRASP1) are consistently higher than the NIST values by  $\sim$1.7 Ryd. The inclusion of Breit and QED effects (GRASP2) affects (lowers) the energies by a maximum of  $\sim$2.1 Ryd, and thus clearly demonstrates the importance of {\em higher} relativistic effects for a heavy ion, such as Kr XXXV. In addition, the orderings have slightly altered in a few instances, see for example the 4/5, 13/14 and 23/24 levels. However, the energy differences for these swapped levels are very small.  In general, the orderings from our {\sc grasp} calculations are nearly the same as  those of NIST, but  energies obtained with the inclusion of  the Breit and QED effects (GRASP2) are consistently lower than the NIST values by  $\sim$0.3 Ryd,  similar to the effect observed for other He-like ions \cite{caxix}. Our FAC1 level energies  are consistently higher by $\sim$0.16 Ryd than the GRASP results, and hence are comparatively in better agreement with the NIST listings. Differences in the {\sc grasp} and {\sc fac} energies arise  mostly by the ways calculations of central potential for radial orbitals and recoupling schemes of angular parts have been performed. Nevertheless, the level orderings from FAC1 are also in general agreement with our calculations from {\sc grasp}, but differ in some instances, particularly for the $n$ = 5 levels. This is mainly because the degeneracy among the levels of the $n$ = 5 configurations is very small. A further inclusion of the 1s6$\ell$ configurations, labelled FAC2 calculations in Table 1, makes no appreciable difference either in the magnitude or orderings of the levels, mainly because the levels of the  1s6$\ell$ configurations lie {\em above} the lowest 49 levels listed in Table 1, and hence do not strongly interact with those. 

Finally, in Table 1 we also list the unpublished energies of Whiteford {\em et al} \cite{icft}, which are obtained from the {\em AutoStructure} ({\sc as}) code of Badnell \cite{as}, and are available at the {\sc apap} (Atomic Processes for Astrophysical Plasmas) website: {\tt {\verb+http://amdpp.phys.strath.ac.uk/UK_APAP/+}}. The {\sc as} energies are higher by up to $\sim$2 Ryd  than the corresponding reference values from NIST, and our theoretical results from the {\sc fac} and {\sc grasp} codes.  This is due to some of the relativistic effects being neglected in the {\sc as} calculations.  More importantly, the level orderings are slightly different particularly for levels of the $n$ = 5 configurations, and the {\sc as}  energies for some of the levels  are non-degenerate. However,  the energy differences among the degenerate levels of a configuration are very small, as noted above. Since the NIST energies are not available for some of the levels, particularly  of the $n$ = 5 configurations,  our energy levels either from the GRASP2 or FAC1 calculations should be adopted in modelling applications. 

\section{Radiative rates}

The absorption oscillator strength ($f_{ij}$) and radiative rate A$_{ji}$ (in s$^{-1}$) for a transition $i \to j$ are related by the following expression:

\begin{equation}
f_{ij} = \frac{mc}{8{\pi}^2{e^2}}{\lambda^2_{ji}} \frac{{\omega}_j}{{\omega}_i} A_{ji}
 = 1.49 \times 10^{-16} \lambda^2_{ji} (\omega_j/\omega_i) A_{ji}
\end{equation}
where $m$ and $e$ are the electron mass and charge, respectively, $c$ is the velocity of light,  $\lambda_{ji}$ is the transition energy/wavelength in $\rm \AA$, and $\omega_i$
and $\omega_j$ are the statistical weights of the lower ($i$) and upper ($j$) levels, respectively. Similarly, the oscillator strength f$_{ij}$ (dimensionless) and the line
strength S (in atomic unit, 1 a.u. = 6.460$\times$10$^{-36}$ cm$^2$ esu$^2$) are related by the  standard equations given below.

\begin{flushleft}
For the electric dipole (E1) transitions 
\end{flushleft} 
\begin{equation}
A_{ji} = \frac{2.0261\times{10^{18}}}{{{\omega}_j}\lambda^3_{ji}} S^{{\rm E1}} \hspace*{0.5 cm} {\rm and} \hspace*{0.5 cm} 
f_{ij} = \frac{303.75}{\lambda_{ji}\omega_i} S^{{\rm E1}}, \\
\end{equation}
\begin{flushleft}
for the magnetic dipole (M1) transitions  
\end{flushleft}
\begin{equation}
A_{ji} = \frac{2.6974\times{10^{13}}}{{{\omega}_j}\lambda^3_{ji}} S^{{\rm M1}} \hspace*{0.5 cm} {\rm and} \hspace*{0.5 cm}
f_{ij} = \frac{4.044\times{10^{-3}}}{\lambda_{ji}\omega_i} S^{{\rm M1}}, \\
\end{equation}
\begin{flushleft}
for the electric quadrupole (E2) transitions 
\end{flushleft}
\begin{equation}
A_{ji} = \frac{1.1199\times{10^{18}}}{{{\omega}_j}\lambda^5_{ji}} S^{{\rm E2}} \hspace*{0.5 cm} {\rm and} \hspace*{0.5 cm}
f_{ij} = \frac{167.89}{\lambda^3_{ji}\omega_i} S^{{\rm E2}}, 
\end{equation}

\begin{flushleft}
and for the magnetic quadrupole (M2) transitions 
\end{flushleft}
\begin{equation}
A_{ji} = \frac{1.4910\times{10^{13}}}{{{\omega}_j}\lambda^5_{ji}} S^{{\rm M2}} \hspace*{0.5 cm} {\rm and} \hspace*{0.5 cm}
f_{ij} = \frac{2.236\times{10^{-3}}}{\lambda^3_{ji}\omega_i} S^{{\rm M2}}. \\
\end{equation}

In Table 2 we present transition energies/wavelengths ($\lambda$, in $\rm \AA$), radiative rates (A$_{ji}$, in s$^{-1}$), oscillator strengths (f$_{ij}$, dimensionless),
and line strengths (S, in a.u.), in length  form only, for all 336 electric dipole (E1) transitions among the 49 levels of Kr XXXV considered here. The {\em indices} used  to represent the lower and upper levels of a transition have already been defined in Table 1. Similarly, there are 391 electric quadrupole (E2), 316  magnetic dipole (M1), and 410
magnetic quadrupole (M2) transitions among the 49 levels. However, for these transitions only the A-values are listed in Table 2, and the corresponding results for f- or S-
values can be easily obtained using Eqs. (1--5).

To assess the accuracy of our A-values, we have performed another calculation with the {\sc fac} code of Gu \cite{fac}.  For all  strong transitions (f $\ge$ 0.01), the A-values from {\sc grasp} and {\sc fac} agree to better than 10\%.  Furthermore, for a majority of the strong E1 transitions (f $\ge$ 0.01) the length and velocity forms in our {\sc grasp} calculations agree to within 10\%. However, the differences are larger for a few  transitions, which are among the degenerate levels of a configuration, such as 4--5, 27--29 and 45--46, because their energy ($\Delta$E) is very small and hence a slight variation in $\Delta$E has a considerable effect on the A-values. For such  transitions  the two forms of the f- value differ by up to two orders of magnitude. Finally, as for the energy levels the effect of additional CI is negligible on the A- values, as results for all transitions agree within 10\% with those obtained with the additional inclusion of the $n$ = 6 configurations. To conclude,  we may state that for almost all strong E1 transitions, our radiative rates are accurate to better than 10\%. However, for the weaker transitions the accuracy is comparatively poorer.

\section{Lifetimes}

The lifetime $\tau$ for a level $j$ is defined as follows:

\begin{equation}  {\tau}_j = \frac{1}{{\sum_{i}^{}} A_{ji}}.  
\end{equation} 
 Since this is a measurable parameter, it provides a check on the accuracy of the calculations. Therefore, in Table 1 we have also listed our calculated lifetimes, which  include the contributions from four types of transitions, i.e. E1, E2, M1, and M2. To our knowledge, no calculations or measurements are available for lifetimes in Kr XXXV with which to compare. However, we hope the present results will be useful for future comparisons and may encourage experimentalists to measure lifetimes, particularly for the level 1s2s $^1$S$_0$ which has comparatively a larger value.

\section{Collision strengths}

Collision strengths ($\Omega$) are related to the more commonly known parameter collision cross section ($\sigma_{ij}$, $\pi{a_0}^2$) by the following relationship:

\begin{equation}
\Omega_{ij}(E) = {k^2_i}\omega_i\sigma_{ij}(E)
\end{equation}
where ${k^2_i}$ is the incident energy of the electron and $\omega_i$ is the statistical weight of the initial state. Results for collisional data are preferred 
in the form of $\Omega$ because it is a symmetric and dimensionless quantity.

For the computation of collision strengths $\Omega$, we have employed the {\em Dirac atomic $R$-matrix code} ({\sc darc}), which includes the relativistic effects in a
systematic way, in both the target description and the scattering model. It is based on the $jj$ coupling scheme, and uses the  Dirac-Coulomb Hamiltonian in the $R$-matrix
approach. The $R$-matrix radius adopted for Kr XXXV is 2.08 au, and 60  continuum orbitals have been included for each channel angular momentum in the expansion of the wavefunction, allowing us to compute $\Omega$ up to an energy of  4000 Ryd,  sufficient to determine values of effective collision strengths $\Upsilon$ (see section 6)  up to T$_e$ = 10$^{8.1}$ K.  The maximum number of channels for a partial wave is 217, and the corresponding size of the Hamiltonian matrix is 13 076. To obtain convergence of  $\Omega$ for all transitions and at all energies, we have included all partial waves with angular momentum $J \le$ 40.5, although a larger number would have been  preferable for the convergence of some allowed transitions, especially at higher energies. However, to account for higher neglected partial waves, we have included a top-up, based on the Coulomb-Bethe approximation \cite{ab} for allowed transitions and geometric series for others.

For illustration, in Figs. 1-3 we show the variation of $\Omega$ with angular momentum $J$ for three transitions of Kr XXXV, namely 2--5 (1s2s $^3$S$_1$ -- 1s2p $^3$P$^o_1$), 2--11 (1s2s $^3$S$_1$ -- 1s3p $^3$P$^o_1$), and 9--12 (1s3p $^3$P$^o_0$ -- 1s3p $^3$P$^o_2$),  and at five energies of 1300, 1800, 2300, 2800 and 3300 Ryd. Values of $\Omega$ have nearly converged for all {\em resonance} transitions (including the allowed ones), and most of the allowed transitions among the higher excited levels,  as shown in Fig. 2 for the 2--11 transition. It is also clear from Fig. 2 that the need to include a larger range of partial waves increases with increasing energy. However, values of $\Omega$ have not converged for those allowed transitions whose $\Delta$E is very small (mainly within the same $n$ complex), as shown for the 2--5 transition in Fig. 1. Similarly, values of $\Omega$ have (almost) converged for all forbidden transitions, including those whose $\Delta$E is very small, such as the 9--12 transition shown in Fig. 3. Therefore, mainly for the allowed transitions within the same $n$ complex, our wide range of partial waves is not sufficient for the convergence of $\Omega$, for which a top-up has been included as mentioned above, and has been found to be appreciable. 

In Table 3 we list our values of $\Omega$ for resonance transitions of Kr XXXV at energies {\em above} thresholds. The  indices used  to represent the levels of a transition have already been defined in Table 1. Unfortunately, no similar data are available for comparison purposes as already stated in section 1. Therefore, in order to make an accuracy assessment for $\Omega$, we have performed another calculation using the {\sc fac} code of Gu \cite{fac}. This code is also fully relativistic, and is based on the well-known and widely-used {\em distorted-wave} (DW) method.  Furthermore, the same CI is included in {\sc fac} as in the calculations from {\sc darc}. Therefore, also included in Table 3 for comparison purposes are the $\Omega$ values from {\sc fac} at a single {\em excited} energy E$_j$, which corresponds to an incident energy of $\sim$ 2800 Ryd. For a majority of transitions the two sets of $\Omega$  generally agree well (within $\sim$ 20\%). However, the differences are larger for a few (particularly weaker) transitions. For example, for 64\% of the Kr XXXV transitions, the values of $\Omega$ agree within 20\% at an energy of $\sim$ 2800 Ryd, and discrepancies for others are mostly within a factor of two, although for some transitions (such as:  19--49, 33--38/41/43/47/49 and 34--36/46/48),  the differences are up to an order of magnitude. However, most of these transitions are weak ($\Omega \le$ 10$^{-6}$) and forbidden, i.e. the values of $\Omega$  have fully converged at {\em all} energies within our adopted range of partial waves in the calculations from the {\sc darc} code. For such weak transitions, values of $\Omega$ from the {\sc fac} code are not assessed to be accurate.  Additionally, for a few transitions values of $\Omega$ from  {\sc fac}  are anomalous, as also noted for other He-like ions and demonstrated in Fig. 6 of  Aggarwal and Keenan \cite{caxix}, \cite{mgxi}. The sudden anomalous behaviour in values of $\Omega$ from  {\sc fac}  is also responsible for the differences noted above for many of the transitions. Such anomalies for some transitions (both allowed and forbidden)  from the  {\sc fac} calculations  arise primarily because of the interpolation and extrapolation techniques employed in the  code, which is designed to generate a large amount of atomic data in a comparatively very short period of time, and without too large loss of accuracy. Similarly, some differences in  $\Omega$ are expected because the DW method generally overestimates the results due to the exclusion of channel coupling. 

As a further comparison between the {\sc darc} and {\sc fac} values of $\Omega$, in Fig. 4 we show the variation of $\Omega$ with energy for three {\em allowed} transitions among the excited levels of Kr XXXV, namely 2--6 (1s2s $^3$S$_1$ - 1s2p $^3$P$^o_2$), 5--14 (1s2p $^3$P$^o_1$ - 1s3d $^3$D$_2$), and 11--24 (1s3p $^3$P$^o_1$ -  1s4d $^3$D$_2$). Also included in this figure are the corresponding results obtained with the {\sc fac} code. For  many  transitions there are no discrepancies between the f- values obtained with the two different codes ({\sc grasp} and {\sc fac}), and therefore the values of $\Omega$ also agree to better than 20\%. However, the values of $\Omega$ obtained with {\sc fac}  differ from our calculations with {\sc darc}, particularly towards the lower end of the energy range. Similar comparisons between the two calculations with {\sc darc} and {\sc fac} are shown in Fig. 5 for three {\em forbidden} transitions of Kr XXXV, namely 2--8 (1s2s $^3$S$_1$ - 1s3s $^3$S$_1$), 2--16 (1s2s $^3$S$_1$ - 1s3d $^3$D$_3$), and 5--11 (1s2p $^3$P$^o_1$ - 1s3p $^3$P$^o_1$). As in the case of  allowed transitions, for these (and many other) forbidden transitions  the agreement between the two calculations improves considerably with increasing energy, but the differences are significant towards the lower end of the energy range.   These anomalies are due to the interpolation and extrapolation techniques employed in the {\sc fac} code, as stated above. Therefore, on the basis of these and other comparisons discussed above, collision strengths from the {\sc fac} code are not assessed to be very accurate, over the entire energy range, for a majority of transitions of Kr XXXV. However, we do not see any apparent deficiency in our calculations for $\Omega$, and estimate our results to be accurate to better than 20\% for a majority of the (strong) transitions. 

\section{Effective collision strengths}

Excitation rates, in addition to energy levels and radiative rates, are required for plasma modelling, and are determined from the collision strengths ($\Omega$). Since the
threshold energy region is dominated by numerous closed-channel (Feshbach) resonances, values of $\Omega$ need to be calculated in a fine energy mesh in order to accurately account for their contribution. Furthermore, in a plasma electrons have a wide distribution of velocities, and therefore values of $\Omega$ are generally averaged over a {\em Maxwellian} distribution as follows:

\begin{equation}
\Upsilon(T_e) = \int_{0}^{\infty} {\Omega}(E) {\rm exp}(-E_j/kT_e) d(E_j/{kT_e}),
\end{equation}
where $k$ is Boltzmann constant, T$_e$ is electron temperature in K, and E$_j$ is the electron energy with respect to the final (excited) state. Once the value of $\Upsilon$ is
known the corresponding results for the excitation q(i,j) and de-excitation q(j,i) rates can be easily obtained from the following equations:

\begin{equation}
q(i,j) = \frac{8.63 \times 10^{-6}}{{\omega_i}{T_e^{1/2}}} \Upsilon {\rm exp}(-E_{ij}/{kT_e}) \hspace*{1.0 cm}{\rm cm^3s^{-1}}
\end{equation}
and
\begin{equation}
q(j,i) = \frac{8.63 \times 10^{-6}}{{\omega_j}{T_e^{1/2}}} \Upsilon \hspace*{1.0 cm}{\rm cm^3 s^{-1}},
\end{equation}
where $\omega_i$ and $\omega_j$ are the statistical weights of the initial ($i$) and final ($j$) states, respectively, and E$_{ij}$ is the transition energy. The contribution of
resonances may enhance the values of $\Upsilon$ over those of the background  collision strengths ($\Omega_B$), especially for the forbidden transitions, by up to an
order of magnitude (or even more) depending on the transition and/or the temperature.  Similarly, values of $\Omega$ need to be calculated over a wide energy range (above
thresholds) in order to obtain convergence of the integral in Eq. (8), as demonstrated in Fig. 7 of Aggarwal and Keenan \cite{ni11a}. 

To properly delineate resonances, we have performed our calculations of $\Omega$ at over $\sim$ 135 000 energies in the thresholds region. Close to thresholds ($\sim$0.1 Ryd above a threshold) the energy mesh is 0.001 Ryd, and away from thresholds is 0.002 Ryd. Thus care has been taken to include as many resonances as possible, and with as fine a resolution as is computationally feasible. The density and importance of resonances can be appreciated from Figs. 7--10, where we plot $\Omega$ as a function of energy in the thresholds region for the four most important transitions of Kr XXXV, namely 1--2 ($z$: 1s$^2$ $^1$S$_0$ -- 1s2s $^3$S$_1$), 1--5 ($y$: 1s$^2$ $^1$S$_0$ -- 1s2p $^3$P$^o_1$),  1--6 ($x$: 1s$^2$ $^1$S$_0$ -- 1s2p $^3$P$^o_2$), and 1--7 ($w$: 1s$^2$ $^1$S$_0$ -- 1s2p $^1$P$^o_1$). For some transitions, such as 1--2, 1--5 and 1--6, the resonances are dense, particularly at energies just above the thresholds. These near threshold resonances affect the values of $\Upsilon$ particularly towards the lower end of the temperature range. Similar dense  resonances have been noted for transitions in other He-like ions \cite{caxix},\cite{mgxi}.

Our calculated values of $\Upsilon$ are listed in Table 4 over a wide temperature range up to 10$^{8.1}$ K, suitable for applications in a variety of plasmas. As stated in section 1, there are only limited results available for comparison purposes. Zhang and Sampson \cite{zs87} have reported values of $\Upsilon$ for transitions among the lowest seven levels of Kr XXXV. In their calculations, they have adopted the Coulomb-Born-exchange method and have also included the contribution of resonances in an approximate way - see \cite{zs87} for details and discussion. Therefore, in Table 5  we compare our results of $\Upsilon$ with theirs at three common temperatures of 1.87$\times$10$^7$, 4.20$\times$10$^7$ and 9.33$\times$10$^7$ K. Generally, the agreement between the two sets of results is  within  about a factor of two, such as  for 4--5 (1s2p $^3$P$^o_1$ -- 1s2s $^1$S$_0$).  This is an `elastic' transition ($\Delta$E $<$ 0.1 Ryd), has a very small f-value ($\sim$10$^{-5}$), and therefore the results of $\Omega$ can be highly variable as also noted for other ions. 

Since the above comparisons are limited to only a few transitions of Kr XXXV, we have also calculated values of $\Upsilon$ from our non-resonant $\Omega$ data obtained with the {\sc fac} code. These calculations are particularly helpful in  providing an estimate of the importance of resonances in the determination of excitation rates. In Table 4  we have included these results from {\sc fac} at the lowest and the highest calculated temperatures for Kr XXXV. At T$_e$ = 10$^{6.7}$ K, our resonances-resolved values of $\Upsilon$ are higher by over 20\% for about 45\% of the transitions of Kr XXXV. Generally, the differences for a majority of the transitions are within a factor of two, but are higher (up to an order of magnitude)  for a few, such as 18--28/29/30/31, 20--29/31, 24--30 and 25--26.  Most of these transitions are forbidden, and the differences are partly due to the contribution of resonances included in our calculations with the {\sc darc} code.  In the case of the most important $w$, $x$, $y$, and $z$ lines, resonances have enhanced the values of $\Upsilon$ by about a  factor of three for the $z$ (1--2: 1s$^2$ $^1$S$_0$ -- 1s2s $^3$S$_1$) transition. A similar comparison at the highest temperature of our calculations, i.e. 10$^{8.1}$ K, indicates that about 40\% of the transitions of Kr XXXV show differences of over 20\% between the {\sc darc} and {\sc fac} values for $\Upsilon$. These differences are generally within a factor of two, but are higher (up to a factor of eight) for a few, such as:  37--45 (1s5d $^3$D$_1$ -- 1s5f $^1$F$_3^o$), 37--47 (1s5d $^3$D$_1$ -- 1s5g $^3$G$_4$), 38--44 (1s5d $^3$D$_2$ -- 1s5f $^3$F$_4^o$)  and 38--48 (1s5d $^3$D$_2$ -- 1s5g $^3$G$_5$).  All these four (and many other) transitions are forbidden, and the $\Upsilon$ from {\sc fac} are {\em higher}, mainly because of the sudden anomalies in the values of $\Omega$, as discussed in section 5. 

Based on the comparisons made above with the  limited published results for transitions in Kr XXXV and  with our  {\sc fac} calculations, we estimate that the accuracy of our values of $\Upsilon$ from the {\sc darc} code listed in Table 4  is better than 20\% for a majority of transitions. A similar conclusion was also made for other He-like ions \cite{caxix}. However,  note that we have  not included the effect of  radiation damping  in our calculations, which may  reduce the importance of resonances in the collision strengths, for some of the transitions, and subsequently in the determination of $\Upsilon$. This effect is important for highly charged ions, because radiative decay rates are large and compete with autoionisation rates  \cite{dpz}. However, earlier $R$-matrix calculations performed by Delahaye {\em et al} \cite{dpz} with the inclusion of radiation damping show that their contribution (i.e. reduction in  values of $\Upsilon$) is appreciable only at lower temperatures (below 10$^6$ K), which may not be important for the plasma modelling  for a highly ionised specie, such as Kr XXXV.  Furthermore, Whiteford {\em et al} \cite{icft} have demonstrated that the resultant uncertainties in the parts of the effective emission coefficients driven by excitation from the ground level for the  four important lines of Ar XVII and Fe XXV (i.e. $w$, $x$, $y$ and $z$), at appropriate temperatures and densities, is under 10\%. Therefore, it is reasonable to conclude that our results of $\Upsilon$ listed in Table 4  will not be significantly affected by the inclusion of radiation damping, and can be confidently applied to the modelling of plasmas. This in fact has been confirmed by the recent calculations of Griffin and Ballance \cite{damp}, who performed similar calculations with the {\sc darc} code, with and  without the inclusion of {\em radiation damping}, and concluded that the difference between the damped and undamped $\Upsilon$ values  averaged over six temperatures (1.25$\times$10$^6$ to 3.12$\times$10$^8$ K)  for 30 resonance transitions is only 4.4\%, and when averaged over 9 temperatures in the range 3.0$\times$10$^6$ to 1.0$\times$10$^9$ K for all 1176 transitions of Kr XXXV, it is only 1.3\%.

In Table 6 we compare our results of $\Upsilon$ with the corresponding {\em radiation damped} values of Griffin and Ballance \cite {damp} for the lowest 30 resonance transitions, which are  common between the two calculations. In general, there is a good agreement between the two independent but similar calculations, although differences are up to $\sim$20\% for some of the transitions, particularly towards the lower end of the temperature range, which is not very important from the applications point of view. For some transitions, such as 1--8/13/14/17, our results are higher, whereas for some others, such as 1--28/29/30/31, those of Griffin and Ballance are higher. Such small differences between any two calculations are not uncommon and mainly arise because of the corresponding differences in the ways the calculations have been performed. For example, Griffin and Ballance calculated $\Omega$ values up to an energy of 2000 Ryd and then extrapolated the data to higher energies in order to calculate corresponding values of $\Upsilon$, whereas our calculations for $\Omega$ are up to E = 4000 Ryd. These differences particularly affect the values of $\Upsilon$ at higher temperatures, and can be better appreciated in Fig. 12 of Aggarwal and Keenan \cite{caxix} and \cite{mgxi}. Similarly, Griffin and Ballance divided their calculations in three parts for three different ranges of partial waves, i.e. $J \le $11.5, 12.5 $\le J \le$ 30.5, and $J \ge$ 31.5, whereas we performed a single calculation for all partial waves. Their procedure sometimes leads to discontinuity at the joining points for some of the transitions, and particularly affect the forbidden transitions. Finally, their energy mesh was comparatively coarser, .i.e. 0.00673 Ryd in comparison to ours of better than 0.002 Ryd, and they had even a broader spacing of 0.648 Ryd for partial waves with $J \ge$ 12.5, whereas we have a uniform energy mesh for {\em all} partial waves. These differences in the energy spacings particularly affect the determination of the contribution of resonances. Considering these differences in the calculation procedures, differences in $\Upsilon$ values noted in Table 6 are fully understandable. However, these comparisons are limited to only $\sim$2.5\% transitions and therefore we discuss below the comparisons among a wider range of transitions.

As for the energy levels listed in Table 1, effective collision strengths for transitions in Kr XXXV are available on the {\sc apap} website: {\tt {\verb+http://amdpp.phys.strath.ac.uk/UK_APAP/+}}. These results were undertaken alongside the earlier calculations of Whiteford {\em et al} \cite{icft} for Ar XVII and Fe XXV. For the calculations, they adopted the standard $R$-matrix code of Berrington {\em et al} \cite{rm}. The electron exchange was included for partial waves up to $J$ = 10.5 but was neglected for higher $J$ values. Furthermore, the calculations were performed in the $LS$ coupling scheme and the corresponding results for $\Omega$ and subsequently $\Upsilon$ for fine-structure transitions were obtained using their intermediate coupling frame transformation ({\sc icft}) method. However, the data obtained by such procedures should qualitatively be comparable with our fully relativistic results from the {\sc darc} code, as has already been demonstrated in several papers -- see for example,  Liang and Badnell \cite{lb}. Furthermore, Whiteford {\em et al} included the effect of radiation damping, which can reduce the contribution of resonances in the determination of $\Upsilon$ values, for some transitions, such as 1s2p $^1$P$^o_1$ -- 1s3s $^1$S$_0$, particularly towards the lower end of the temperature range -- see their Fig. 4.  However, at temperatures relevant to plasma modelling applications for Kr XXXV, the effect of radiation damping is negligible as discussed above.

As stated above, the two independent sets of results from {\sc darc} and Whiteford {\em et al} \cite{icft} should be broadly comparable, although relativistic effects are very important for a heavy ion considered in this paper, and noted earlier for energy levels in section 2. However, differences between the two datasets  are considerable (over 20\%) for many transitions, and throughout the temperature range of the calculations. For example, at the lowest common temperature of 2.45$\times$10$^5$ K, the two sets of $\Upsilon$ differ by over 20\% for $\sim$40\% of the transitions. For a majority of transitions, these differences are within a factor of three. For some transitions our results are higher, but for most others the reverse is true. Of particular note are those transitions for which the differences are over two orders of magnitude, and the $\Upsilon$ values of Whiteford {\em et al} are invariably higher. Examples of such transitions are: 13--14/16/17, 14--16/17 and 16--17. Most of these transitions belong to the degenerate levels of a state/configuration, and hence have very small transition energies, as seen in Table 1. Such transitions may be referred to as `elastic' and are always problematic to calculate -- see for example Hamada {\em et al} \cite{hydro} for hydrogenic ions which show similar problems.  To demonstrate the differences between the two calculations,  in Fig. 10 we compare values of $\Upsilon$  for three transitions of Kr XXXV, namely  13--14 (1s3d $^3$D$_{1}$ --  1s3d $^3$D$_{2}$),  14--16 (1s3d $^3$D$_{2}$ -- 1s3d $^3$D$_{3}$), and 29--31 (1s4f $^3$F$^o_{3}$ -- 1s4f $^1$F$^o_{3}$). These differences in the $\Upsilon$ values are {\em not} due to any resonances, but arise from the limitation of the approach adopted by Whiteford {\em et al} \cite{icft}, as recently discussed and demonstrated by Bautista et al. \cite{feiii}.  Similar large differences are also observed with the calculations of Whiteford {\em et al} \cite{ar16} for transitions of Li-like ions, as discussed and demonstrated by Aggarwal and Keenan \cite{lia},\cite{lib}.
The problem in the $R$-matrix code adopted by Whiteford {\em et al} \cite{icft},\cite{ar16} has been identified and rectified by Liang and Badnell \cite{lb}. However, since the calculations of Whiteford {\em et al} \cite{icft} were performed more than a decade ago, limitations in their data for $\Upsilon$ remain, and a reexamination of their results is desirable.

The differences between our results of $\Upsilon$ from {\sc darc} and those of Whiteford {\em et al} \cite{icft} are not confined to lower temperatures, but are also found at higher temperatures. At the highest common temperature of 1.22$\times$10$^8$ K, the two sets of $\Upsilon$ differ by over 20\% for $\sim$15\% of the transitions of Kr XXXV. Hence, there is comparatively a better convergence of the results at higher temperatures. However, for almost all  such transitions, the $\Upsilon$ values of Whiteford {\em et al} are higher. The differences for most transitions are within a factor of two, but for some are up to an order of magnitude,  such as: 14--16/17, 19--43/45, 29--31 and 33--43/45. All these transitions are {\em forbidden} and $\Omega$ for these have {\em converged} within our adopted partial waves range, as discussed in section 5. To demonstrate the differences, in Fig. 11 we compare our results of $\Upsilon$ from {\sc darc} with those of Whiteford {\em et al} \cite{icft} for three transitions of Kr XXXV, namely  14--17 (1s3d $^3$D$_{1}$ --  1s3d $^1$D$_{2}$), 46--48 (1s5g $^3$G$_3$ -- 1s5g $^3$G$_5$) and  47--49 (1s5g $^3$G$_{4}$ -- 1s5g $^1$G$_4$). Since the collision strengths of Whiteford {\em et al} are  overestimated for many transitions,  the corresponding results for $\Upsilon$ are affected throughout the entire temperature range.  

To conclude, we may state that the $\Upsilon$ results of Whiteford {\em et al} \cite{icft} should not be as accurate as those presented here for the reasons given above. Based on several comparisons as well as with our experience with other He-like ions \cite{caxix}, \cite{mgxi} and  \cite{arxvii}, we estimate that the accuracy of our $\Upsilon$ results from the {\sc darc} code listed in Table 4 to be better than 20\% for a majority of transitions. 

\section{Conclusions}

In this paper we have presented results for energy levels and  radiative rates for four types of transitions (E1, E2, M1, and M2) among the lowest 49 levels of Kr XXXV belonging to the $n \le$ 5 configurations. Additionally, lifetimes of all the calculated levels have been reported, although  no measurements or other theoretical results are available for comparison.  However, based on a variety of comparisons among various calculations with the {\sc grasp} and {\sc fac} codes, our results for radiative rates, oscillator strengths, line strengths, and lifetimes are judged to be accurate to better than 20\% for a majority of the strong transitions (levels).  Additionally, we have considered a large number of partial waves to achieve convergence of $\Omega$ at all energies, included a wide energy range to accurately calculate  values of $\Upsilon$ up to T$_e$ = 10$^{8.1}$ K, and resolved resonances in a fine energy mesh to account for their contributions. Hence we see no apparent deficiency in our reported results for collision strengths and effective collision strengths, and  estimate their accuracy to be better than 20\% for most transitions.  However, the present data for effective collision strengths for transitions involving the levels of the $n$ = 5 configurations may  be improved somewhat by the inclusion of the levels of the $n$ = 6 configurations. We believe the present set of {\em complete} results for radiative and excitation rates for {\em all} transitions of Kr XXXV will be highly useful for the modelling of a variety of plasmas. 




\clearpage

%
\begin{figure*}
 \includegraphics[scale=0.70,angle=-90]{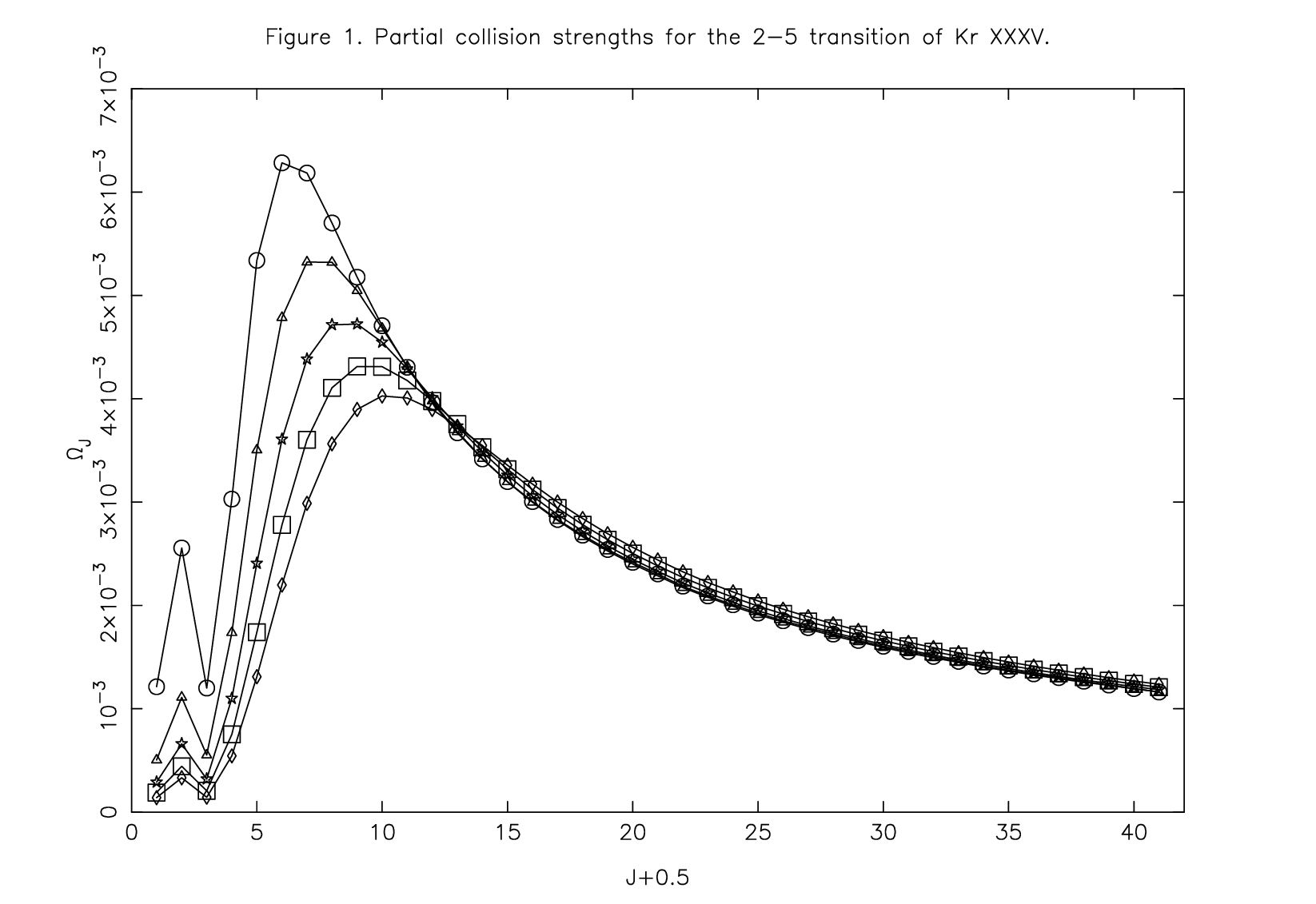}
\caption{Partial collision strengths for the 1s2s $^3$S$_1$ - 1s2p $^3$P$^o_1$ (2--5) transition of Kr XXXV, 
at five energies of: 1300 Ryd (circles), 1800 Ryd (triangles), 2300 Ryd (stars), 2800 Ryd (squares), and 3300 Ryd (diamonds).}
\label{fig:1}       
\end{figure*}

\begin{figure*}
\includegraphics[scale=0.70,angle=-90]{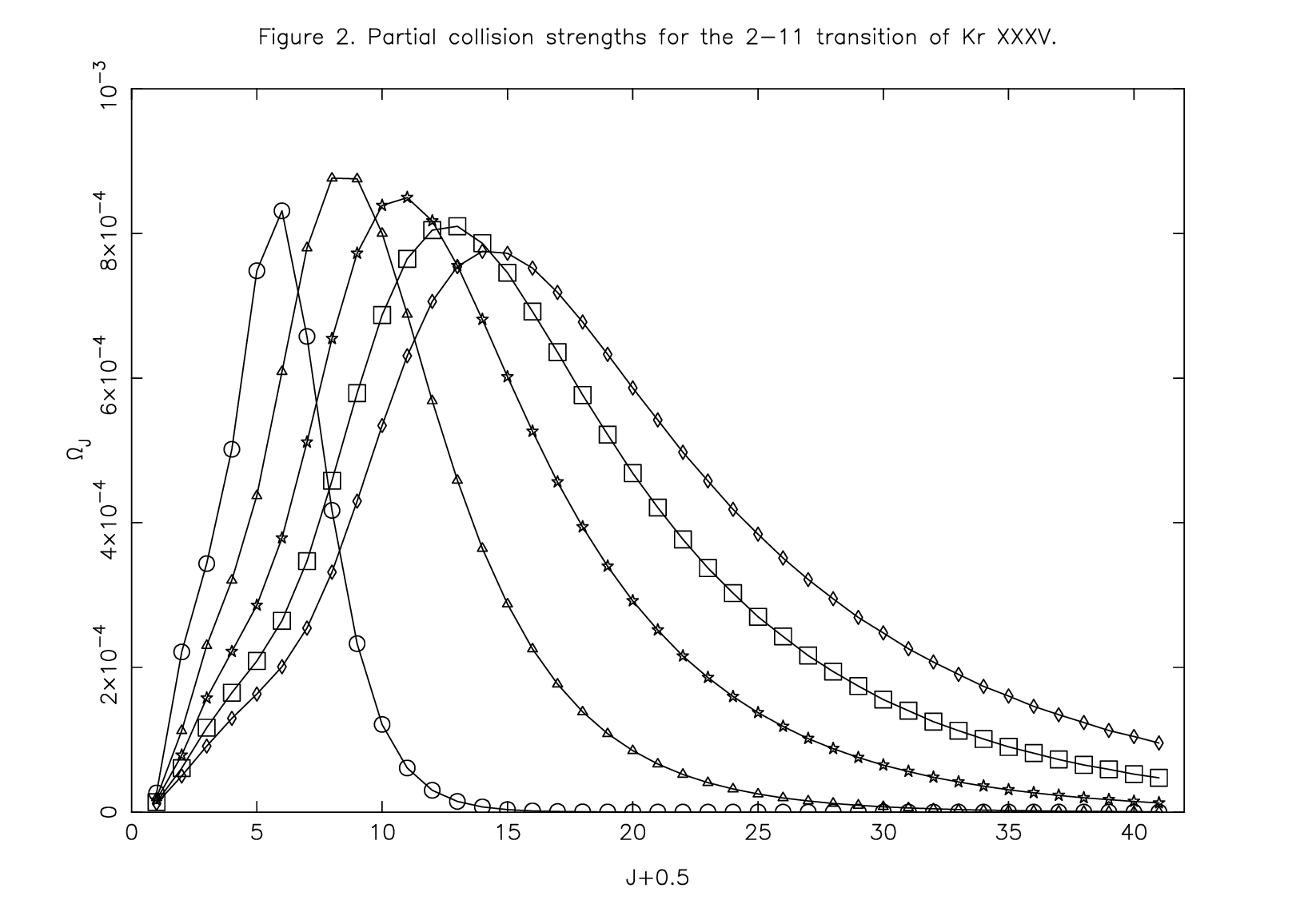}
\caption{Partial collision strengths for the 1s2s $^3$S$_1$ - 1s3p $^3$P$^o_1$ (2--11) transition of Kr XXXV, 
at five energies of: 1300 Ryd (circles), 1800 Ryd (triangles), 2300 Ryd (stars), 2800 Ryd (squares), and 3300 Ryd (diamonds).}
\label{fig:2}       
\end{figure*}
\clearpage
%

\begin{figure*}
\includegraphics[scale=0.70,angle=-90]{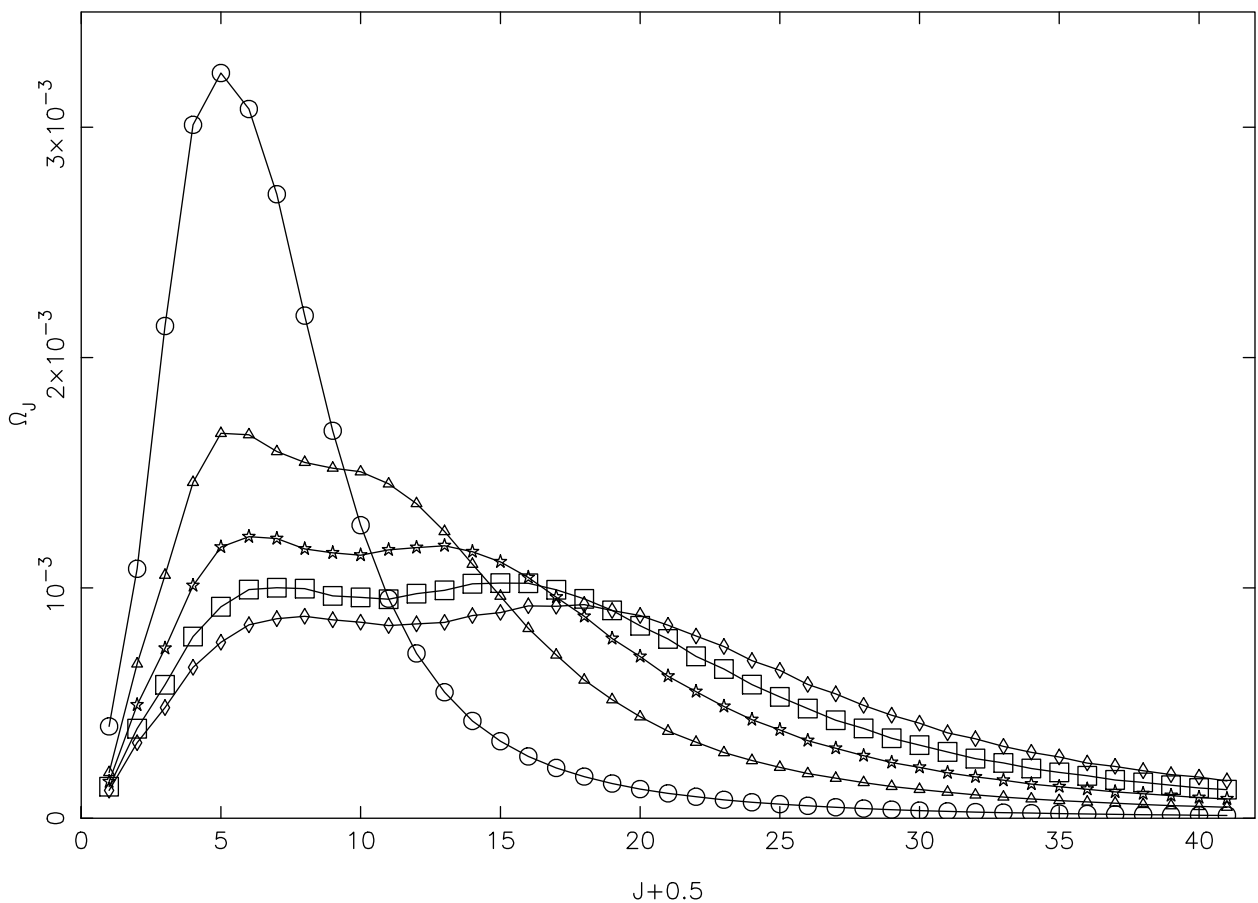}
\caption{Partial collision strengths for the 1s3p $^3$P$^o_0$ - 1s3p $^3$P$^o_2$ (9--12) transition of Kr XXXV, 
at five energies of: 1300 Ryd (circles), 1800 Ryd (triangles), 2300 Ryd (stars), 2800 Ryd (squares), and 3300 Ryd (diamonds).}
\label{fig:3}       
\end{figure*}

\clearpage

\begin{figure*}
\includegraphics[scale=0.70,angle=-90]{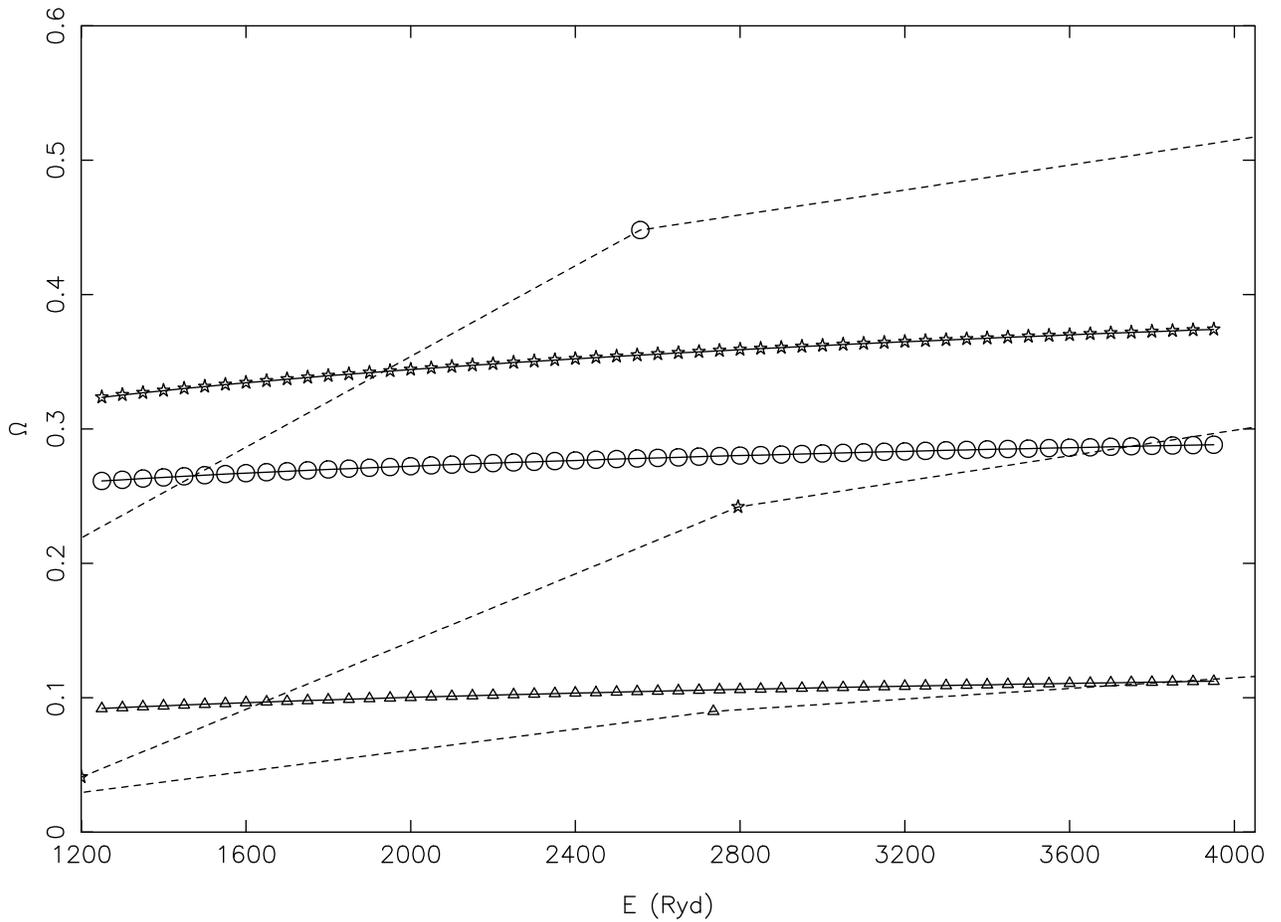}
\caption{Comparison of collision strengths from our calculations from {\sc darc} (continuous curves) and {\sc fac} (broken curves) for the  2--6 (circles: 1s2s $^3$S$_1$ - 1s2p $^3$P$^o_2$), 5--14 (triangles : 1s2p $^3$P$^o_1$ - 1s3d $^3$D$_2$), and 11--24 (stars: 1s3p $^3$P$^o_1$ -  1s4d $^3$D$_2$) allowed transitions of Kr XXXV.}
\label{fig:4}       
\end{figure*}

\clearpage

\begin{figure*}
\includegraphics[scale=0.70,angle=-90]{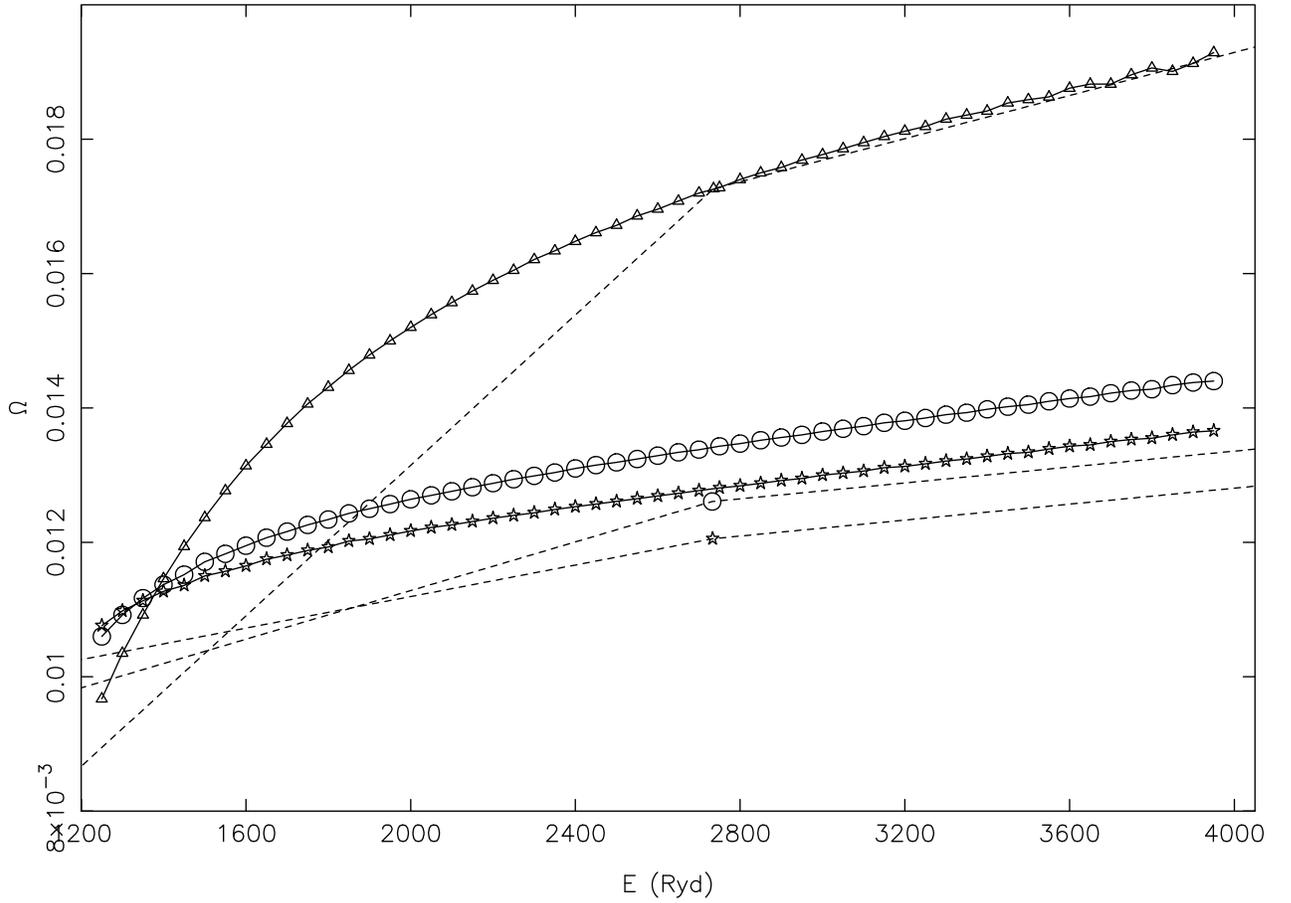}
\caption{Comparison of collision strengths from our calculations from {\sc darc} (continuous curves) and {\sc fac} (broken curves) for the  2--8 (circles: 1s2s $^3$S$_1$ - 1s3s
$^3$S$_1$), 2--16 (triangles: 1s2s $^3$S$_1$ - 1s3d $^3$D$_3$), and 5--11 (stars: 1s2p $^3$P$^o_1$ - 1s3p $^3$P$^o_1$) forbidden transitions of Kr XXXV.}
\label{fig:5}       
\end{figure*}
\clearpage

\clearpage

\begin{figure*}
\includegraphics[scale=0.70,angle=90]{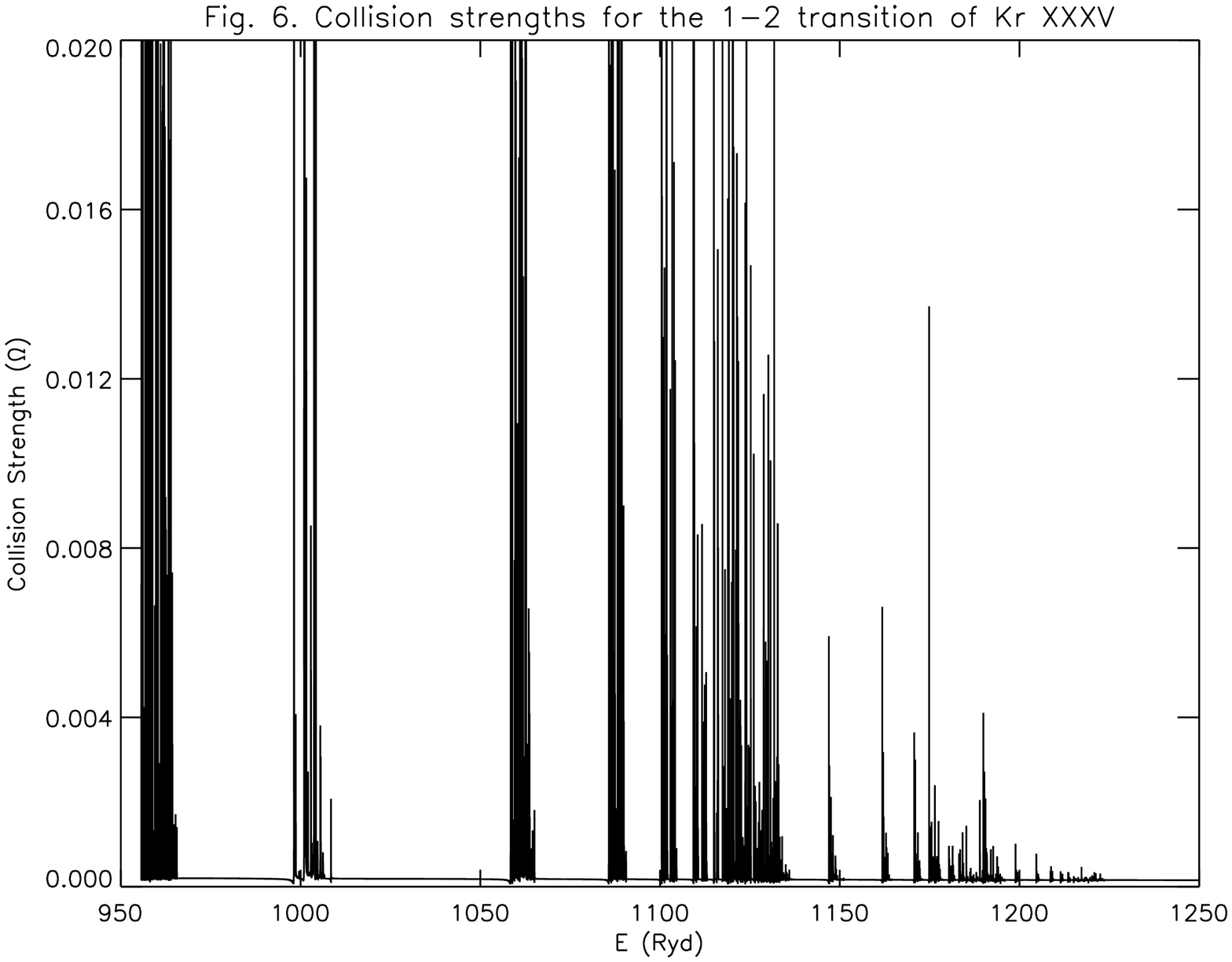}
\caption{Collision strengths for the 1s$^2$ $^1$S$_0$ - 1s2s $^3$S$_1$   (1--2) transition of Kr XXXV.}
\label{fig:7}       
\end{figure*}

\begin{figure*}
\includegraphics[scale=0.70,angle=90]{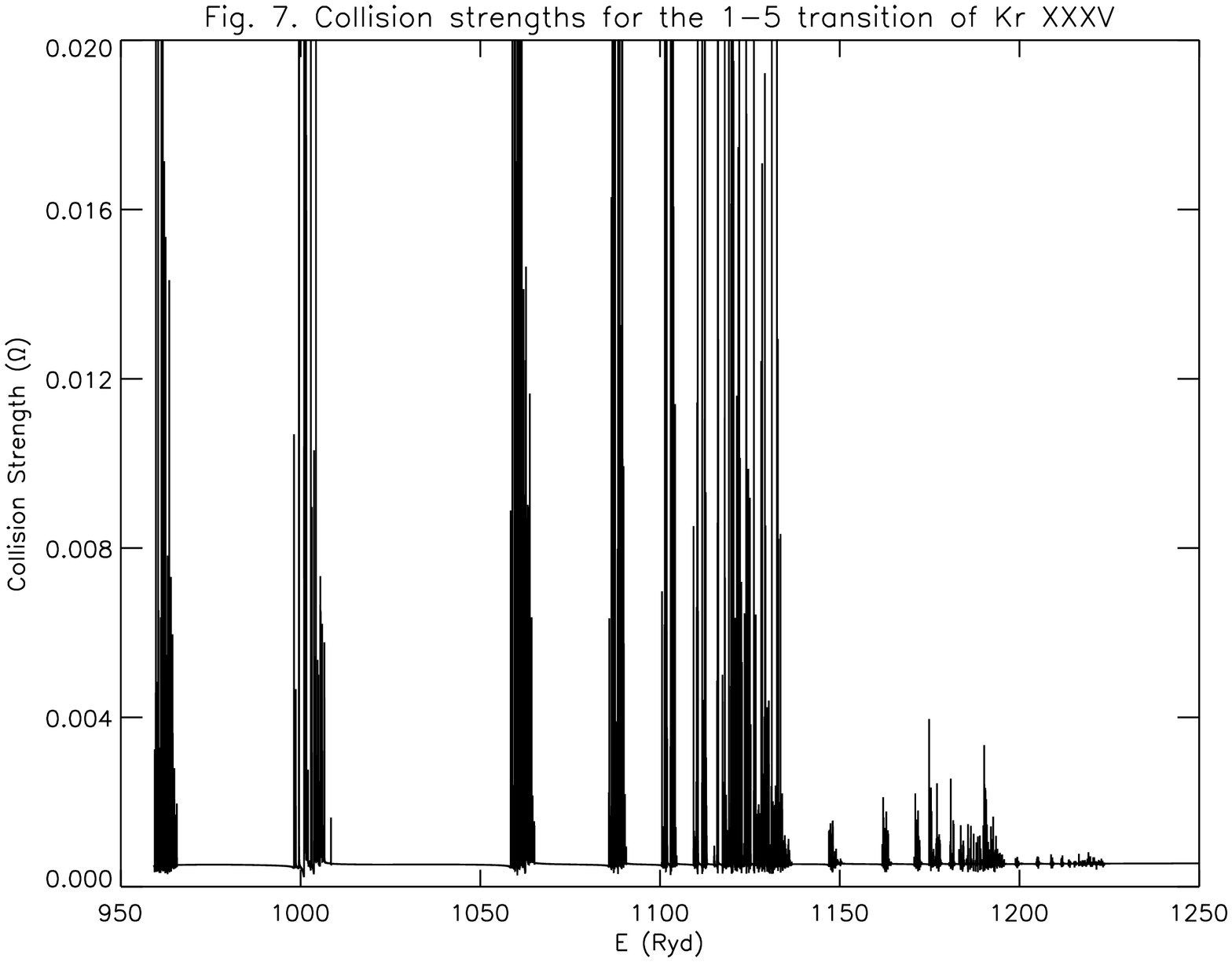}
\caption{Collision strengths for the 1s$^2$ $^1$S$_0$ - 1s2p $^3$P$^o_1$ (1--5) transition of Kr XXXV.}
\label{fig:8}       
\end{figure*}

\begin{figure*}
\includegraphics[scale=0.70,angle=90]{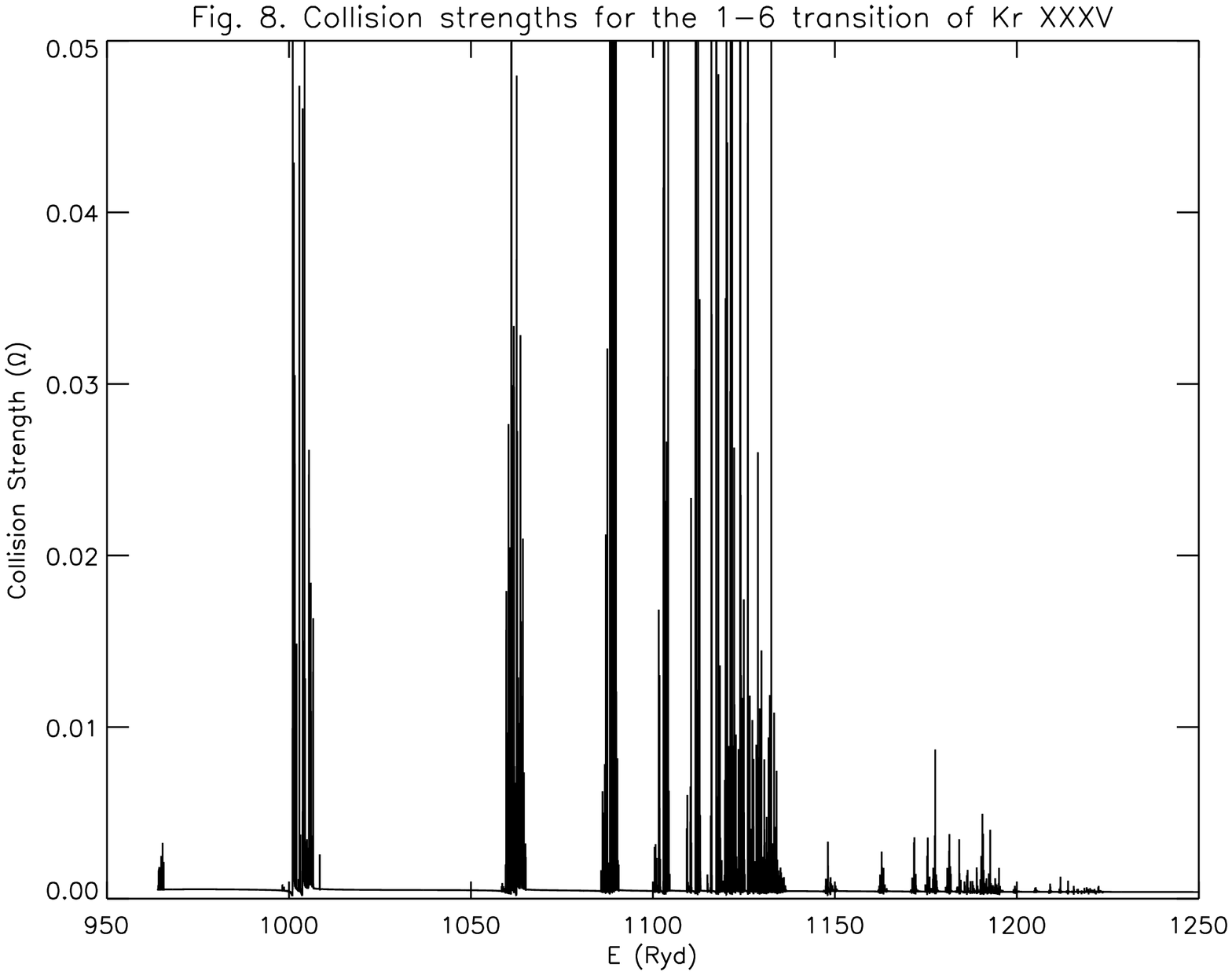}
\caption{Collision strengths for the 1s$^2$ $^1$S$_0$ - 1s2p $^3$P$^o_2$ (1--6) transition of Kr XXXV.}
\label{fig:9}       
\end{figure*}

\begin{figure*}
\includegraphics[scale=0.70,angle=90]{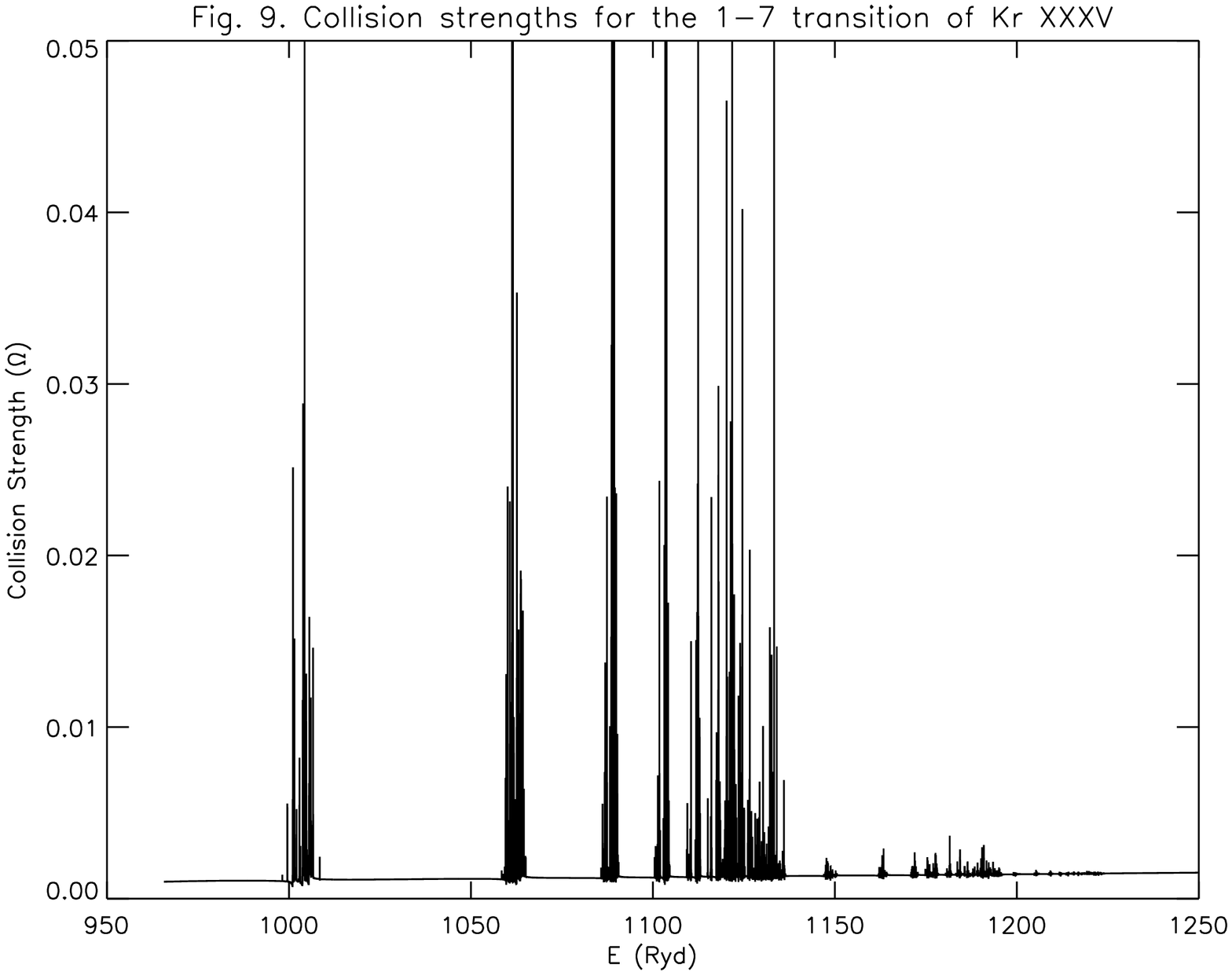}
\caption{Collision strengths for the 1s$^2$ $^1$S$_0$ - 1s2p $^1$P$^o_1$ (1--7) transition of Kr XXXV.}
\label{fig:10}       
\end{figure*}

\setcounter{figure} {9}
\begin{figure*}
\includegraphics[scale=0.70,angle=-90]{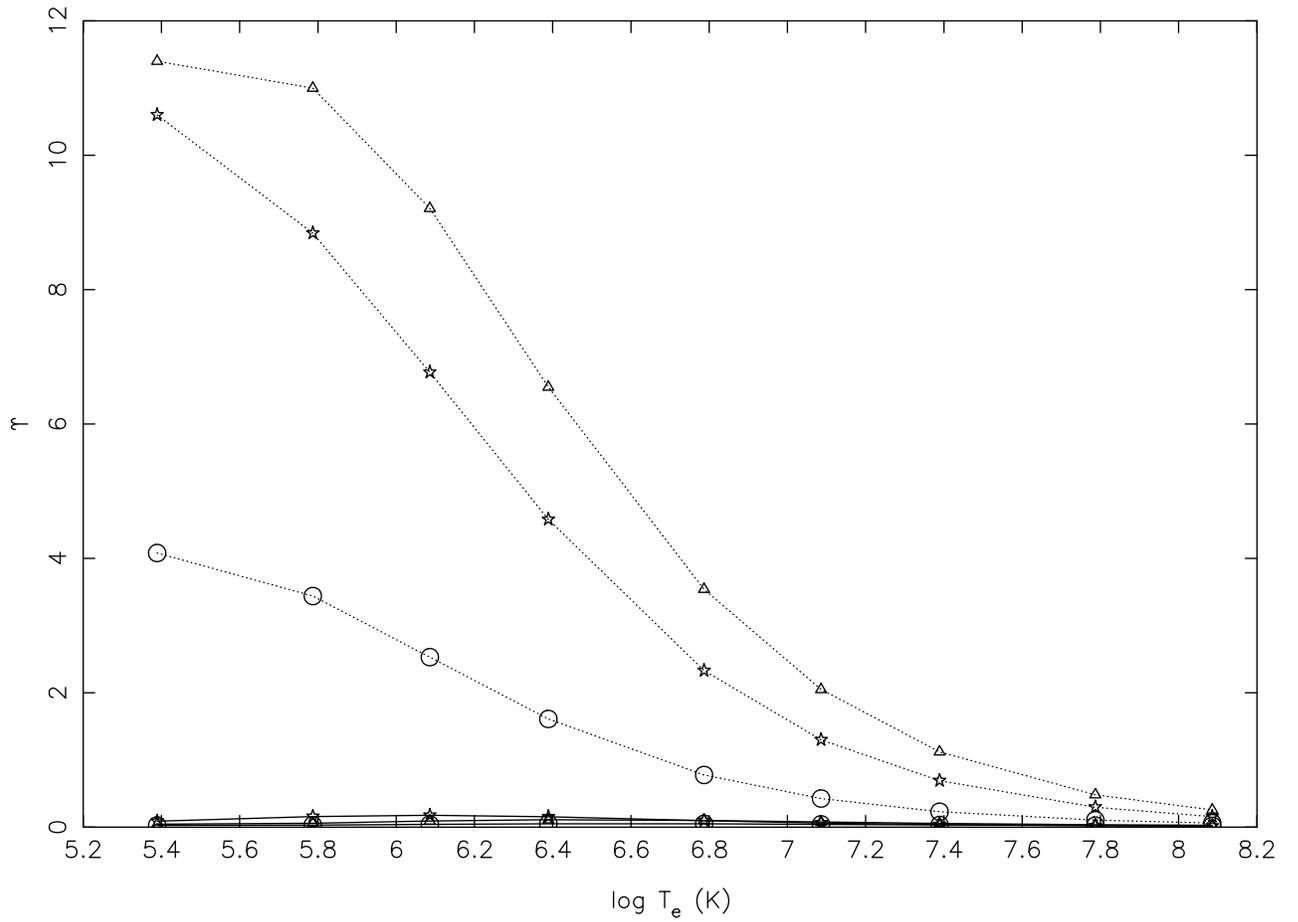}
\caption{Comparison of effective collision strengths for the 13--14 (circles: 1s3d $^3$D$_{1}$ --  1s3d $^3$D$_{2}$), 
14--16 (triangles: 1s3d $^3$D$_{2}$ -- 1s3d $^3$D$_{3}$), and 29--31 (stars: 1s4f $^3$F$^o_{3}$ -- 1s4f $^1$F$^o_{3}$)
 transitions of Kr XXXV. Continuous and dotted curves are from the present {\sc darc} and earlier $R$- matrix codes \cite{icft}, respectively.}
\end{figure*}

\setcounter{figure} {10}
\begin{figure*}
\includegraphics[scale=0.70,angle=-90]{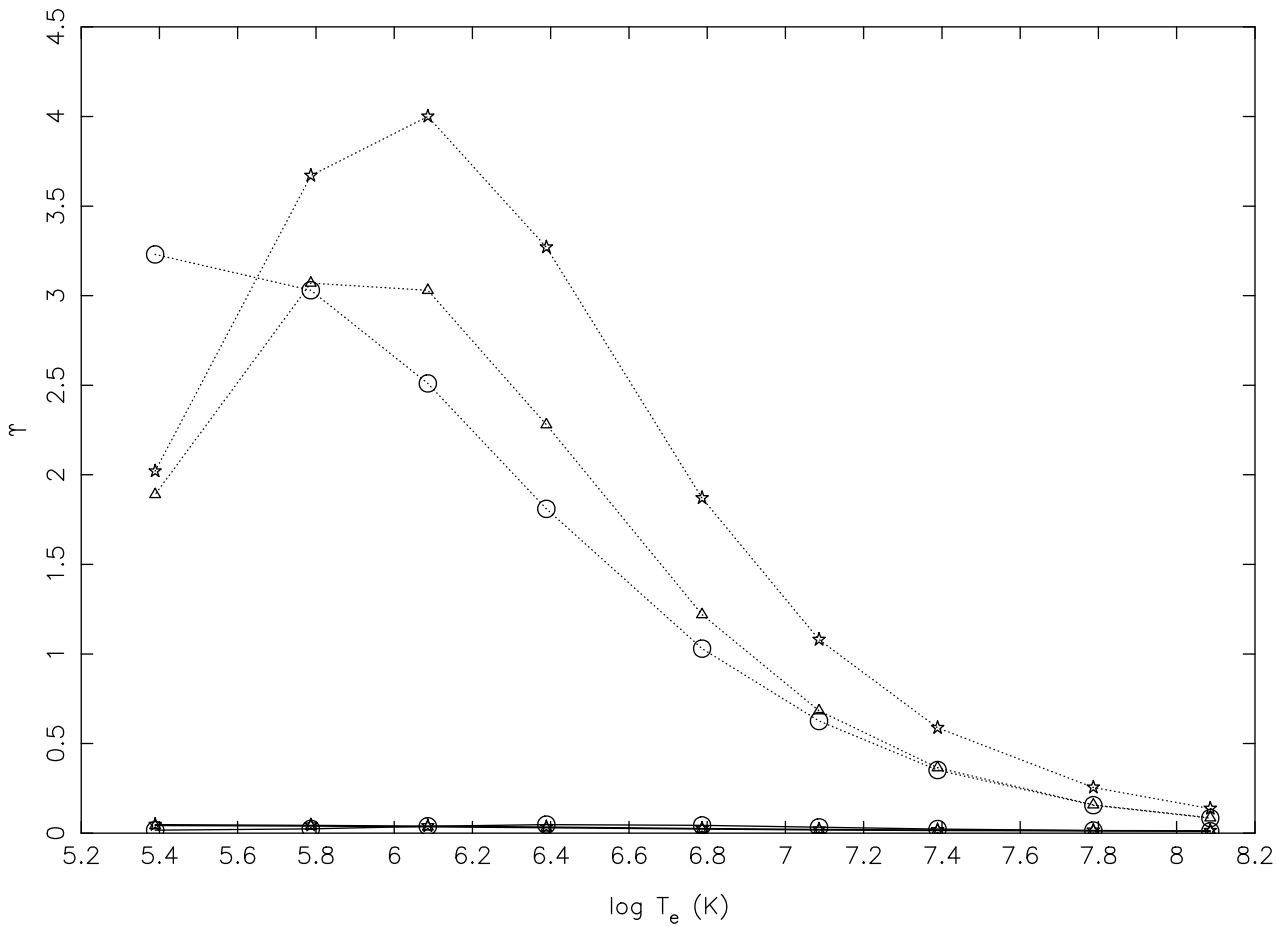}
\caption{Comparison of effective collision strengths for the 14--17 (circles: 1s3d $^3$D$_{2}$ --  1s3d $^1$D$_{2}$), 
46--48 (1s5g $^3$G$_3$ -- 1s5g $^3$G$_5$), and 47--49 (stars: 1s5g $^3$G$_{4}$ -- 1s5g $^1$G$_{4}$)
 transitions of Kr XXXV. Continuous and dotted curves are from the present {\sc darc} and earlier $R$- matrix codes \cite{icft}, respectively.}
\end{figure*}


\clearpage

\begin{table*} 
\caption{Energy levels (in Ryd) of Kr XXXV and their lifetimes ($\tau$, s). $a{\pm}b \equiv a{\times}$10$^{{\pm}b}$.} 
\begin{tabular}{rllrrrrrrrr} \hline
Index  & \multicolumn{2}{c}{Configuration/Level} & NIST  & GRASP1 & GRASP2      &   FAC1          &    FAC2  &  AS          &    $\tau$ (s) \\	 
\hline
1   &  1s$^2$    &  $^1$S$_0  $    &     0.00000   &     0.00000    &	   0.00000  &     0.00000  &     0.00000  &  	0.00000   &    ........   \\
2   &  1s2s	 &  $^3$S$_1  $    &   953.96146   &   955.65820    &	 953.67554  &   953.82037  &   953.82007  &   955.92340   &    1.750-10   \\
3   &  1s2p	 &  $^3$P$_0^o$    &   957.21396   &   958.59088    &	 956.93842  &   957.11481  &   957.11475  &   958.70093   &    1.563-09   \\
4   &  1s2s	 &  $^1$S$_0  $    &   957.43640   &   958.94348    &	 957.20795  &   957.36169  &   957.36133  &   959.41003   &    6.766-04   \\
5   &  1s2p	 &  $^3$P$_1^o$    &   957.40569   &   959.18024    &	 957.13043  &   957.31805  &   957.31787  &   959.30499   &    2.535-15   \\
6   &  1s2p	 &  $^3$P$_2^o$    &   962.16505   &   963.95660    &	 961.88519  &   962.06219  &   962.06207  &   964.03949   &    9.634-12   \\
7   &  1s2p	 &  $^1$P$_1^o$    &   963.89846   &   965.65979    &	 963.63324  &   963.82947  &   963.82935  &   965.81641   &    6.538-16   \\
8   &  1s3s	 &  $^3$S$_1  $    &  1131.57220   &  1133.35706    &	1131.28918  &  1131.45544  &  1131.45520  &  1133.63391   &    8.530-14   \\
9   &  1s3p	 &  $^3$P$_0^o$    &  1132.46620   &  1134.16785    &	1132.18726  &  1132.35498  &  1132.35498  &  1134.38318   &    2.858-14   \\
10  &  1s3s	 &  $^1$S$_0  $    &  1132.50400   &  1134.22363    &	1132.22754  &  1132.38708  &  1132.38672  &  1134.48926   &    8.531-14   \\
11  &  1s3p	 &  $^3$P$_1^o$    &  1132.51900   &  1134.32898    &	1132.24121  &  1132.41150  &  1132.41150  &  1134.53625   &    6.888-15   \\
12  &  1s3p	 &  $^3$P$_2^o$    &  1133.93790   &  1135.75427    &	1133.65955  &  1133.82922  &  1133.82910  &  1135.88745   &    3.044-14   \\
13  &  1s3d	 &  $^3$D$_1  $    &  1134.39900   &  1136.19482    &	1134.11792  &  1134.27722  &  1134.27722  &  1136.31445   &    9.990-15   \\
14  &  1s3d	 &  $^3$D$_2  $    &  1134.37350   &  1136.20605    &	1134.09302  &  1134.25281  &  1134.25281  &  1136.32727   &    9.942-15   \\
15  &  1s3p	 &  $^1$P$_1^o$    &  1134.41360   &  1136.21631    &	1134.13977  &  1134.31030  &  1134.31018  &  1136.34875   &    2.206-15   \\
16  &  1s3d	 &  $^3$D$_3  $    &  1134.86650   &  1136.69421    &	1134.58447  &  1134.74365  &  1134.74365  &  1136.82690   &    1.023-14   \\
17  &  1s3d	 &  $^1$D$_2  $    &  1134.90660   &  1136.71252    &	1134.62476  &  1134.78479  &  1134.78479  &  1136.84705   &    1.018-14   \\
18  &  1s4s	 &  $^3$S$_1  $    &  1193.11010   &  1194.92407    &	1192.83704  &  1192.99402  &  1192.99377  &  1195.15540   &    1.259-13   \\
19  &  1s4p	 &  $^3$P$_0^o$    &  1193.48190   &  1195.25732    &	1193.20520  &  1193.35474  &  1193.35461  &  1195.45972   &    4.916-14   \\
20  &  1s4s	 &  $^1$S$_0  $    &  1193.49370   &  1195.27576    &	1193.21899  &  1193.37097  &  1193.37085  &  1195.48157   &    1.194-13   \\
21  &  1s4p	 &  $^3$P$_1^o$    &  1193.50650   &  1195.32336    &	1193.22766  &  1193.37817  &  1193.37817  &  1195.51843   &    1.530-14   \\
22  &  1s4p	 &  $^3$P$_2^o$    &  1194.10340   &  1195.92480    &	1193.82617  &  1193.97717  &  1193.97705  &  1196.06384   &    5.183-14   \\
23  &  1s4d	 &  $^3$D$_1  $    &  1194.29700   &  1196.10571    &	1194.01501  &  1194.17737  &  1194.17737  &  1196.23499   &    2.321-14   \\
24  &  1s4d	 &  $^3$D$_2  $    &  1194.28650   &  1196.11169    &	1194.00586  &  1194.16846  &  1194.16846  &  1196.24158   &    2.308-14   \\
25  &  1s4p	 &  $^1$P$_1^o$    &  1194.30300   &  1196.11414    &	1194.02380  &  1194.17419  &  1194.17407  &  1196.24646   &    5.230-15   \\
26  &  1s4d	 &  $^3$D$_3  $    &  1194.49520   &  1196.31665    &	1194.21216  &  1194.37390  &  1194.37390  &  1196.44812   &    2.384-14   \\
27  &  1s4d	 &  $^1$D$_2  $    &  1194.51160   &  1196.32666    &	1194.23120  &  1194.39343  &  1194.39343  &  1196.45544   &    2.364-14   \\
28  &  1s4f	 &  $^3$F$_2^o$    &  		   &  1196.32764    &	1194.23071  &  1194.38342  &  1194.38354  &  1196.45886   &    4.768-14   \\
29  &  1s4f	 &  $^3$F$_3^o$    &  		   &  1196.32764    &	1194.22388  &  1194.37671  &  1194.37671  &  1196.45532   &    4.767-14   \\
30  &  1s4f	 &  $^3$F$_4^o$    &  		   &  1196.43225    &	1194.32898  &  1194.48169  &  1194.48169  &  1196.56287   &    4.804-14   \\
31  &  1s4f	 &  $^1$F$_3^o$    &  		   &  1196.43237    &	1194.33423  &  1194.48694  &  1194.48694  &  1196.56299   &    4.807-14   \\
32  &  1s5s	 &  $^3$S$_1  $    &  1221.42130   &  1223.24036    &	1221.14648  &  1221.29895  &  1221.29871  &  1223.44763   &    2.031-13   \\
33  &  1s5p	 &  $^3$P$_0^o$    &  		   &  1223.40881    &	1221.33228  &  1221.47986  &  1221.47986  &  1223.59973   &    8.482-14   \\
34  &  1s5s	 &  $^1$S$_0  $    &  1221.61640   &  1223.41772    &	1221.33960  &  1221.48828  &  1221.48804  &  1223.59961   &    1.725-13   \\
35  &  1s5p	 &  $^3$P$_1^o$    &  1221.62270   &  1223.44214    &	1221.34375  &  1221.49182  &  1221.49182  &  1223.62720   &    2.996-14   \\
36  &  1s5p	 &  $^3$P$_2^o$    &  1221.92800   &  1223.74976    &	1221.64990  &  1221.79822  &  1221.79810  &  1223.89001   &    8.918-14   \\
37  &  1s5d	 &  $^3$D$_1  $    &  		   &  1223.84106    &	1221.74548  &  1221.90564  &  1221.90564  &  1223.97449   &    4.468-14   \\
38  &  1s5d	 &  $^3$D$_2  $    &  		   &  1223.84448    &	1221.74097  &  1221.90137  &  1221.90137  &  1223.97791   &    4.403-14   \\
39  &  1s5p	 &  $^1$P$_1^o$    &  1222.03100   &  1223.84558    &	1221.75012  &  1221.89758  &  1221.89746  &  1223.97864   &    1.060-14   \\
40  &  1s5d	 &  $^3$D$_3  $    &  		   &  1223.94922    &	1221.84631  &  1222.00610  &  1222.00610  &  1224.08069   &    4.604-14   \\
41  &  1s5d	 &  $^1$D$_2  $    &  		   &  1223.95471    &	1221.85657  &  1222.01672  &  1222.01672  &  1224.08484   &    4.473-14   \\
42  &  1s5f	 &  $^3$F$_2^o$    &  		   &  1223.95532    &	1221.85632  &  1222.00867  &  1222.00867  &  1224.08655   &    9.195-14   \\
43  &  1s5f	 &  $^3$F$_3^o$    &  		   &  1223.95532    &	1221.85291  &  1222.00525  &  1222.00525  &  1224.08484   &    9.198-14   \\
44  &  1s5f	 &  $^3$F$_4^o$    &  		   &  1224.00891    &	1221.90674  &  1222.05908  &  1222.05908  &  1224.13916   &    9.273-14   \\
45  &  1s5f	 &  $^1$F$_3^o$    &  		   &  1224.00891    &	1221.90942  &  1222.06177  &  1222.06177  &  1224.13916   &    9.277-14   \\
46  &  1s5g	 &  $^3$G$_3  $    &  		   &  1224.00903    &	1221.90918  &  1222.06128  &  1222.06128  &  1224.13794   &    1.555-13   \\
47  &  1s5g	 &  $^3$G$_4  $    &  		   &  1224.00903    &	1221.90710  &  1222.05920  &  1222.05920  &  1224.13794   &    1.555-13   \\
48  &  1s5g	 &  $^3$G$_5  $    &  		   &  1224.04102    &	1221.93933  &  1222.09143  &  1222.09143  &  1224.17090   &    1.560-13   \\
49  &  1s5g	 &  $^1$G$_4  $    &  		   &  1224.04102    &	1221.94092  &  1222.09302  &  1222.09302  &  1224.17090   &    1.561-13   \\
\hline 													     
\end{tabular}      
			      
\begin{flushleft}
{\small
NIST: {\tt http://nist.gov/pml/data/asd.cfm} \\
GRASP1: Energies from the {\sc grasp} code with 49 level calculations {\em without} Breit and QED effects \\
GRASP2: Energies from the {\sc grasp} code with 49 level calculations {\em with} Breit and QED effects \\
FAC1: Energies from the {\sc fac} code with 49 level calculations \\
FAC2: Energies from the {\sc fac} code with 71 level calculations \\
AS: Energies from the {\sc as} code with 49 level calculations \\
}
\end{flushleft}
\end{table*} 

\clearpage

\setcounter{table}{2} 

\begin{table*}                                                                                                
\caption{Collision strengths for transitions in  Kr XXXV. ($a{\pm}b \equiv$ $a\times$10$^{{\pm}b}$).}     
\begin{tabular}{rrlllllll}                                                                                   
\hline                                                                                                        
\hline                                                                                                        
\multicolumn{2}{c}{Transition} & \multicolumn{6}{c}{Energy (Ryd)}\\                                           
\hline                                                                                                        
   $i$ & $j$ &    1500 &   2000 &   2500 &  3000  & 3500  &  FAC$^a$ \\                                         
\hline                                                                                                        
   1 &  2 &  1.234$-$4 &  8.438$-$5 &  6.348$-$5 &  4.832$-$5 &  3.739$-$5 &  5.626$-$5 \\
  1 &  3 &  6.580$-$5 &  3.991$-$5 &  2.671$-$5 &  1.901$-$5 &  1.407$-$5 &  2.314$-$5 \\
  1 &  4 &  5.523$-$4 &  6.240$-$4 &  6.814$-$4 &  7.261$-$4 &  7.644$-$4 &  5.744$-$4 \\
  1 &  5 &  5.933$-$4 &  6.989$-$4 &  8.130$-$4 &  9.230$-$4 &  1.027$-$3 &  9.387$-$4 \\
  1 &  6 &  2.862$-$4 &  1.704$-$4 &  1.124$-$4 &  7.888$-$5 &  5.770$-$5 &  1.068$-$4 \\
  1 &  7 &  1.917$-$3 &  2.629$-$3 &  3.255$-$3 &  3.807$-$3 &  4.308$-$3 &  3.328$-$3 \\
  1 &  8 &  3.825$-$5 &  2.593$-$5 &  1.875$-$5 &  1.405$-$5 &  1.107$-$5 &  1.482$-$5 \\
  1 &  9 &  2.143$-$5 &  1.285$-$5 &  8.350$-$6 &  5.885$-$6 &  4.352$-$6 &  6.242$-$6 \\
  1 & 10 &  9.769$-$5 &  1.164$-$4 &  1.287$-$4 &  1.392$-$4 &  1.487$-$4 &  1.120$-$4 \\
  1 & 11 &  1.192$-$4 &  1.315$-$4 &  1.485$-$4 &  1.667$-$4 &  1.845$-$4 &  1.766$-$4 \\
  1 & 12 &  9.532$-$5 &  5.601$-$5 &  3.584$-$5 &  2.491$-$5 &  1.818$-$5 &  2.920$-$5 \\
  1 & 13 &  7.256$-$6 &  3.608$-$6 &  2.077$-$6 &  1.319$-$6 &  8.987$-$7 &  1.509$-$6 \\
  1 & 14 &  1.580$-$5 &  1.698$-$5 &  1.956$-$5 &  2.223$-$5 &  2.469$-$5 &  2.198$-$5 \\
  1 & 15 &  3.031$-$4 &  4.374$-$4 &  5.526$-$4 &  6.536$-$4 &  7.443$-$4 &  6.022$-$4 \\
  1 & 16 &  1.505$-$5 &  7.323$-$6 &  4.145$-$6 &  2.603$-$6 &  1.758$-$6 &  3.376$-$6 \\
  1 & 17 &  1.813$-$5 &  2.457$-$5 &  3.083$-$5 &  3.632$-$5 &  4.104$-$5 &  3.336$-$5 \\
  1 & 18 &  1.666$-$5 &  1.105$-$5 &  8.069$-$6 &  6.002$-$6 &  4.682$-$6 &  6.001$-$6 \\
  1 & 19 &  9.417$-$6 &  5.550$-$6 &  3.619$-$6 &  2.524$-$6 &  1.871$-$6 &  2.555$-$6 \\
  1 & 20 &  3.501$-$5 &  4.211$-$5 &  4.745$-$5 &  5.134$-$5 &  5.520$-$5 &  4.138$-$5 \\
  1 & 21 &  4.571$-$5 &  4.864$-$5 &  5.440$-$5 &  6.064$-$5 &  6.706$-$5 &  6.450$-$5 \\
  1 & 22 &  4.224$-$5 &  2.440$-$5 &  1.565$-$5 &  1.076$-$5 &  7.875$-$6 &  1.200$-$5 \\
  1 & 23 &  4.249$-$6 &  2.074$-$6 &  1.195$-$6 &  7.513$-$7 &  5.121$-$7 &  8.184$-$7 \\
  1 & 24 &  8.024$-$6 &  7.971$-$6 &  9.004$-$6 &  1.020$-$5 &  1.135$-$5 &  1.030$-$5 \\
  1 & 25 &  1.048$-$4 &  1.538$-$4 &  1.964$-$4 &  2.332$-$4 &  2.666$-$4 &  2.169$-$4 \\
  1 & 26 &  8.854$-$6 &  4.228$-$6 &  2.393$-$6 &  1.487$-$6 &  1.004$-$6 &  1.835$-$6 \\
  1 & 27 &  8.618$-$6 &  1.139$-$5 &  1.437$-$5 &  1.708$-$5 &  1.943$-$5 &  1.604$-$5 \\
  1 & 28 &  2.445$-$7 &  9.833$-$8 &  4.895$-$8 &  2.794$-$8 &  1.754$-$8 &  3.202$-$8 \\
  1 & 29 &  3.402$-$7 &  2.468$-$7 &  2.406$-$7 &  2.502$-$7 &  2.623$-$7 &  2.479$-$7 \\
  1 & 30 &  3.940$-$7 &  1.553$-$7 &  7.637$-$8 &  4.320$-$8 &  2.697$-$8 &  5.597$-$8 \\
  1 & 31 &  3.323$-$7 &  2.844$-$7 &  3.032$-$7 &  3.281$-$7 &  3.508$-$7 &  3.056$-$7 \\
  1 & 32 &  8.547$-$6 &  5.552$-$6 &  4.071$-$6 &  3.119$-$6 &  2.310$-$6 &  2.988$-$6 \\
  1 & 33 &  4.934$-$6 &  2.863$-$6 &  1.857$-$6 &  1.303$-$6 &  9.546$-$7 &  1.285$-$6 \\
  1 & 34 &  1.711$-$5 &  2.046$-$5 &  2.305$-$5 &  2.504$-$5 &  2.691$-$5 &  2.013$-$5 \\
  1 & 35 &  2.273$-$5 &  2.368$-$5 &  2.632$-$5 &  2.929$-$5 &  3.231$-$5 &  3.105$-$5 \\
  1 & 36 &  2.224$-$5 &  1.264$-$5 &  8.068$-$6 &  5.574$-$6 &  4.036$-$6 &  6.034$-$6 \\
  1 & 37 &  2.479$-$6 &  1.195$-$6 &  6.858$-$7 &  4.313$-$7 &  2.928$-$7 &  4.587$-$7 \\
  1 & 38 &  4.437$-$6 &  4.215$-$6 &  4.702$-$6 &  5.312$-$6 &  5.907$-$6 &  5.409$-$6 \\
  1 & 39 &  4.979$-$5 &  7.343$-$5 &  9.411$-$5 &  1.119$-$4 &  1.280$-$4 &  1.040$-$4 \\
  1 & 40 &  5.178$-$6 &  2.441$-$6 &  1.377$-$6 &  8.555$-$7 &  5.754$-$7 &  1.029$-$6 \\
  1 & 41 &  4.609$-$6 &  5.958$-$6 &  7.526$-$6 &  8.975$-$6 &  1.024$-$5 &  8.521$-$6 \\
  1 & 42 &  2.063$-$7 &  8.201$-$8 &  4.069$-$8 &  2.318$-$8 &  1.454$-$8 &  2.601$-$8 \\
  1 & 43 &  2.820$-$7 &  1.923$-$7 &  1.842$-$7 &  1.920$-$7 &  2.031$-$7 &  1.906$-$7 \\
  1 & 44 &  3.329$-$7 &  1.298$-$7 &  6.357$-$8 &  3.586$-$8 &  2.237$-$8 &  4.550$-$8 \\
  1 & 45 &  2.733$-$7 &  2.191$-$7 &  2.310$-$7 &  2.518$-$7 &  2.720$-$7 &  2.353$-$7 \\
  1 & 46 &  4.283$-$9 &  1.317$-$9 &  5.39$-$10 &  2.67$-$10 &  1.52$-$10 &  3.17$-$10 \\
  1 & 47 &  5.723$-$9 &  2.526$-$9 &  1.840$-$9 &  1.774$-$9 &  1.847$-$9 &  1.553$-$9 \\
  1 & 48 &  6.040$-$9 &  1.823$-$9 &  7.36$-$10 &  3.62$-$10 &  2.05$-$10 &  4.87$-$10 \\
  1 & 49 &  5.714$-$9 &  2.742$-$9 &  2.158$-$9 &  2.169$-$9 &  2.300$-$9 &  1.790$-$9 \\
\hline                                                                                                        
\end{tabular}     
\begin{flushleft}
{\small
$^a$: E $\sim$ 2800 Ryd  \\
}
\end{flushleft}                                                                                            
\end{table*}                                                                                                  
                                                               
\clearpage              
\setcounter{table}{4}                                                                                         
\begin{table*}                                                                                                
\caption{Comparison of $\Upsilon$ values for transitions of  Kr XXXV. ($a{\pm}b \equiv$ $a\times$10$^{{\pm}b}$).}            
\begin{tabular}{rrllllll}                                                                                    
\hline                                                                                                        
\hline 
\multicolumn{2}{c}{T$_e$ (K)} &  \multicolumn{2}{c}{1.87$\times$10$^7$} &   \multicolumn{2}{c}{4.20$\times$10$^7$} &   
\multicolumn{2}{c}{9.33$\times$10$^7$}  \\ 
\hline                                                                                                                                                                                                      
  $i$ & $j$ &   DARC & ZS &   DARC & ZS &   DARC & ZS  \\					       
\hline                                          					      
    1  &  2  &  2.80$-$4 &  2.00$-$4 & 2.21$-$4  &  1.69$-$4 & 1.64$-$4   &  1.32$-$4  \\
    1  &  3  &  1.33$-$4 &  1.06$-$4 & 1.10$-$4  &  8.93$-$5 & 8.26$-$5   &  6.83$-$5  \\
    1  &  4  &  5.05$-$4 &  4.13$-$4 & 5.18$-$4  &  4.31$-$4 & 5.52$-$4   &  4.63$-$4  \\
    1  &  5  &  6.01$-$4 &  5.34$-$4 & 6.00$-$4  &  5.46$-$4 & 6.33$-$4   &  5.92$-$4  \\
    1  &  6  &  5.99$-$4 &  5.43$-$4 & 4.96$-$4  &  4.54$-$4 & 3.68$-$4   &  3.46$-$4  \\
    1  &  7  &  1.28$-$3 &  1.28$-$3 & 1.50$-$3  &  1.50$-$3 & 1.92$-$3   &  1.93$-$3  \\
    2  &  3  &  4.74$-$2 &  6.32$-$2 & 5.04$-$2  &  7.20$-$2 & 5.32$-$2   &  8.22$-$2  \\
    2  &  4  &  2.22$-$3 &  1.60$-$3 & 1.64$-$3  &  1.22$-$3 & 1.11$-$3   &  8.57$-$4  \\
    2  &  5  &  1.16$-$1 &  1.49$-$1 & 1.24$-$1  &  1.70$-$1 & 1.31$-$1   &  1.94$-$1  \\
    2  &  6  &  2.27$-$1 &  2.47$-$1 & 2.51$-$1  &  2.80$-$1 & 2.71$-$1   &  3.21$-$1  \\
    2  &  7  &  2.89$-$2 &  2.81$-$2 & 3.03$-$2  &  3.08$-$2 & 3.16$-$2   &  3.44$-$2  \\
    3  &  4  &  4.70$-$4 &  3.18$-$4 & 3.24$-$4  &  2.20$-$4 & 1.93$-$4   &  1.35$-$4  \\
    3  &  5  &  5.28$-$3 &  4.02$-$3 & 4.03$-$3  &  3.10$-$3 & 2.73$-$3   &  2.19$-$3  \\
    3  &  6  &  6.16$-$3 &  4.80$-$3 & 5.26$-$3  &  4.20$-$3 & 4.52$-$3   &  3.74$-$3  \\
    3  &  7  &  3.65$-$3 &  2.68$-$3 & 2.47$-$3  &  1.80$-$3 & 1.45$-$3   &  1.07$-$3  \\
    4  &  5  &  2.52$-$2 &  5.74$-$2 & 2.68$-$2  &  6.47$-$2 & 2.85$-$2   &  7.21$-$2  \\
    4  &  6  &  2.62$-$3 &  1.85$-$3 & 1.79$-$3  &  1.24$-$3 & 1.05$-$3   &  7.42$-$4  \\
    4  &  7  &  1.13$-$1 &  1.24$-$1 & 1.24$-$1  &  1.42$-$1 & 1.33$-$1   &  1.64$-$1  \\
    5  &  6  &  2.06$-$2 &  1.58$-$2 & 1.62$-$2  &  1.26$-$2 & 1.22$-$2   &  9.90$-$3  \\
    5  &  7  &  1.08$-$2 &  7.87$-$3 & 8.49$-$3  &  6.29$-$3 & 6.45$-$3   &  5.02$-$3  \\
    6  &  7  &  1.70$-$2 &  1.19$-$2 & 1.27$-$2  &  8.91$-$3 & 8.55$-$3   &  6.30$-$3  \\
\hline                                                                                                        
\end{tabular}  

\begin{flushleft}
{\small
DARC: Present calculations from the DARC code \\
ZS:   Calculations of Zhang and Sampson \cite{zs87} \\

}
\end{flushleft}
                                                                                               
\end{table*}                                                      


\clearpage  

\setcounter{table}{5}                                                                                         
\begin{table*}                                                                                                
\caption{Comparison of $\Upsilon$ values for resonance transitions of  Kr XXXV. ($a{\pm}b \equiv$ $a\times$10$^{{\pm}b}$).}            
\begin{tabular}{rrllllllll}                                                                                    
\hline                                                                                                        
\hline 
\multicolumn{2}{c}{Transition/T$_e$ (K)} &  \multicolumn{2}{c}{5.0$\times$10$^6$} &   \multicolumn{2}{c}{1.0$\times$10$^7$} &   
\multicolumn{2}{c}{5.0$\times$10$^7$}  &   \multicolumn{2}{c}{1.0$\times$10$^8$} \\ 
\hline                                                                                                                                                                                                      
  $i$ & $j$ &   AK & GB &    AK & GB  &   AK & GB  &   AK & GB   \\					       
\hline                                          					      
    1  &    2  &    3.524$-$4  &  4.19$-$4  &    3.178$-$4  &	3.37$-$4  &   2.086$-$4  &  2.03$-$4  &    1.596$-$4  &   1.55$-$4 \\	   
    1  &    3  &    1.460$-$4  &  1.48$-$4  &    1.428$-$4  &	1.36$-$4  &   1.043$-$4  &  9.84$-$5  &    8.017$-$5  &   7.73$-$5 \\	   
    1  &    4  &    4.990$-$4  &  4.80$-$4  &    5.008$-$4  &	4.77$-$4  &   5.242$-$4  &  5.09$-$4  &    5.548$-$4  &   5.41$-$4 \\	   
    1  &    5  &    5.889$-$4  &  5.58$-$4  &    6.004$-$4  &	5.63$-$4  &   6.038$-$4  &  5.80$-$4  &    6.373$-$4  &   6.26$-$4 \\	   
    1  &    6  &    6.264$-$4  &  5.88$-$4  &    6.363$-$4  &	5.80$-$4  &   4.688$-$4  &  4.33$-$4  &    3.569$-$4  &   3.38$-$4 \\	   
    1  &    7  &    1.120$-$3  &  1.10$-$3  &    1.192$-$3  &	1.16$-$3  &   1.578$-$3  &  1.55$-$3  &    1.965$-$3  &   1.94$-$3 \\	   
    1  &    8  &    8.969$-$5  &  7.30$-$5  &    7.799$-$5  &	6.70$-$5  &   4.984$-$5  &  4.69$-$5  &    3.916$-$5  &   3.76$-$5 \\	   
    1  &    9  &    3.978$-$5  &  3.69$-$5  &    3.703$-$5  &	3.49$-$5  &   2.615$-$5  &  2.56$-$5  &    2.043$-$5  &   2.06$-$5 \\	   
    1  &   10  &    1.015$-$4  &  8.82$-$5  &    9.858$-$5  &	8.90$-$5  &   9.886$-$5  &  9.55$-$5  &    1.054$-$4  &   1.01$-$4 \\	   
    1  &   11  &    1.448$-$4  &  1.33$-$4  &    1.394$-$4  &	1.31$-$4  &   1.276$-$4  &  1.25$-$4  &    1.300$-$4  &   1.30$-$4 \\	   
    1  &   12  &    1.773$-$4  &  1.65$-$4  &    1.656$-$4  &	1.57$-$4  &   1.167$-$4  &  1.15$-$4  &    9.068$-$5  &   9.14$-$5 \\	   
    1  &   13  &    2.684$-$5  &  2.10$-$5  &    2.335$-$5  &	1.89$-$5  &   1.228$-$5  &  1.10$-$5  &    8.538$-$6  &   7.89$-$6 \\	   
    1  &   14  &    4.067$-$5  &  3.24$-$5  &    3.529$-$5  &	2.93$-$5  &   2.214$-$5  &  2.04$-$5  &    1.995$-$5  &   1.92$-$5 \\	   
    1  &   15  &    2.206$-$4  &  2.12$-$4  &    2.273$-$4  &	2.22$-$4  &   2.902$-$4  &  2.87$-$4  &    3.569$-$4  &   3.49$-$4 \\	   
    1  &   16  &    4.862$-$5  &  4.28$-$5  &    4.228$-$5  &	3.85$-$5  &   2.345$-$5  &  2.25$-$5  &    1.659$-$5  &   1.62$-$5 \\	   
    1  &   17  &    3.417$-$5  &  2.85$-$5  &    2.982$-$5  &	2.62$-$5  &   2.226$-$5  &  2.13$-$5  &    2.335$-$5  &   2.28$-$5 \\	   
    1  &   18  &    3.472$-$5  &  2.98$-$5  &    2.856$-$5  &	2.59$-$5  &   1.905$-$5  &  1.85$-$5  &    1.529$-$5  &   1.52$-$5 \\	   
    1  &   19  &    1.560$-$5  &  1.55$-$5  &    1.415$-$5  &	1.41$-$5  &   1.027$-$5  &  1.03$-$5  &    8.120$-$6  &   8.24$-$6 \\	   
    1  &   20  &    3.495$-$5  &  3.31$-$5  &    3.342$-$5  &	3.23$-$5  &   3.531$-$5  &  3.48$-$5  &    3.819$-$5  &   3.70$-$5 \\	   
    1  &   21  &    5.395$-$5  &  5.31$-$5  &    5.092$-$5  &	5.05$-$5  &   4.777$-$5  &  4.77$-$5  &    4.866$-$5  &   4.91$-$5 \\	   
    1  &   22  &    7.621$-$5  &  6.84$-$5  &    6.800$-$5  &	6.29$-$5  &   4.723$-$5  &  4.60$-$5  &    3.685$-$5  &   3.68$-$5 \\	   
    1  &   23  &    1.031$-$5  &  1.05$-$5  &    8.659$-$6  &	8.79$-$6  &   5.236$-$6  &  5.28$-$6  &    3.850$-$6  &   3.89$-$6 \\	   
    1  &   24  &    1.580$-$5  &  1.57$-$5  &    1.320$-$5  &	1.31$-$5  &   9.497$-$6  &  9.49$-$6  &    8.953$-$6  &   9.02$-$6 \\	   
    1  &   25  &    8.413$-$5  &  8.07$-$5  &    8.428$-$5  &	8.21$-$5  &   1.061$-$4  &  1.05$-$4  &    1.299$-$4  &   1.26$-$4 \\	   
    1  &   26  &    2.066$-$5  &  2.13$-$5  &    1.756$-$5  &	1.79$-$5  &   1.080$-$5  &  1.09$-$5  &    7.939$-$6  &   8.04$-$6 \\	   
    1  &   27  &    1.306$-$5  &  1.37$-$5  &    1.114$-$5  &	1.15$-$5  &   9.726$-$6  &  9.79$-$6  &    1.069$-$5  &   1.07$-$5 \\	   
    1  &   28  &    2.795$-$6  &  3.45$-$6  &    1.806$-$6  &	2.20$-$6  &   6.055$-$7  &  6.96$-$7  &    3.707$-$7  &   4.18$-$7 \\	   
    1  &   29  &    3.531$-$6  &  4.32$-$6  &    2.264$-$6  &	2.74$-$6  &   7.920$-$7  &  9.04$-$7  &    5.378$-$7  &   6.02$-$7 \\	   
    1  &   30  &    3.743$-$6  &  4.52$-$6  &    2.454$-$6  &	2.93$-$6  &   8.693$-$7  &  9.83$-$7  &    5.422$-$7  &   6.02$-$7 \\	   
    1  &   31  &    3.423$-$6  &  4.35$-$6  &    2.168$-$6  &	2.73$-$6  &   7.634$-$7  &  8.94$-$7  &    5.388$-$7  &   6.10$-$7 \\	   

\hline                                                                                                        
\end{tabular}  

\begin{flushleft}
{\small
AK: Present calculations from the DARC code without radiation damping  \\
GB: Calculations of Griffin and Ballance \cite{damp} from the DARC code  with radiation damping\\

}
\end{flushleft}
                                                                                               
\end{table*}                                                                                                                                                               
 
\end{document}